\PassOptionsToPackage{dvipsnames}{xcolor}

\documentclass[%
reprint,
amsmath,amssymb,
aps,
prl,
]{revtex4-2}

\usepackage[dvipsnames]{xcolor}
\usepackage{graphicx}
\usepackage{dcolumn}
\usepackage{bm}
\usepackage{lipsum}
\usepackage{natbib}
\usepackage{hyperref}
\usepackage{subfiles}
\usepackage{times}
\usepackage{tikz}
\usepackage{amsmath}
\usepackage[margin=2cm]{geometry}
\usetikzlibrary{arrows.meta, calc}

\newcommand{\norm}[1]{\left\lVert#1\right\rVert}

\newcommand{\dd}{\mathrm{d}}
\newcommand{\e}{\mathrm{e}}

\newcommand{\sign}{\operatorname{sign}}

\begin{document}

\title{Spontaneous and Collective Learning in a Minimal Agent-Environment Model}

\author{Sebastiano Ariosto}
\thanks{Both authors contributed equally.}
\author{Jérôme Garnier-Brun}
\thanks{Both authors contributed equally.}
\author{Luca Saglietti}
\author{Davide Straziota}
\thanks{This work was conducted prior to the author's employment at Amazon and is unrelated to his Amazon role.}
\affiliation{%
Bocconi Institute for Data Science and Analytics (BIDSA) \& Department of Computing Sciences, Bocconi University, Milan, Italy
}

\date{\today}

\begin{abstract}
How can information be extracted from data without explicit guidance or rewards? We investigate this question in a minimal setting where a classifying agent is exposed to a stream of structured data produced by a generative environment, and evolves by seeking consistency of its own labels in time. We find that imposing a simplicity bias on the classification rule can drive the dynamics toward label-coherent steady states, which we coin \emph{functional replicators}. Remarkably, these persistent labeling rules align with the latent structure of the data. Using analytical tools from statistical mechanics, we characterize this spontaneous learning phase transition. Extending the analysis to a collective of agents that pool labels from one another, we show that interaction reshapes the learning phase boundary, in some regimes enabling spontaneous learning that no isolated agent can achieve, while suppressing it in others. Our minimal framework thus opens a route to decentralized learning through label exchange alone, requiring no access to the internal weights of other agents.
\end{abstract}

\maketitle

A striking feature of nature is the spontaneous emergence of organization from simple interaction rules, ranging from crystallization~\cite{kushner1969self} and molecular self-assembly~\cite{stradner2004equilibrium,higgs2015rna} to collective motion in active matter~\cite{ball1999self,whitesides2002self,vicsek1995novel}. This question is central to living systems, where persistence is achieved not through equilibrium but through \textit{self-replication}. 
Facing this observation, the field of Artificial Life has long explored how complexity can arise and persist \textit{in silico} from minimal substrates~\cite{gardner1970mathematical,langton1986studying, fontana1994arrival,holland1995hidden,chan2018lenia}, and how adaptive or even “intelligent” behavior can manifest spontaneously~\cite{ackley1991interactions,littman1996simulations, wong2023roles}. 
More recently, \citet{y2024computational} demonstrated that in a population of random executable programs without predefined objectives, computation can undergo a sudden surge and stabilize in the form of self-replicating code. 
While evocative, these ideas remain to be formalized in a systematic theoretical framework for learning systems.

Over the past decades, statistical mechanics has elucidated how simplified neural network models solve high-dimensional tasks~\cite{engel2001statistical,advani2013statistical,gabrie2023neural,cui2025high}, yet such analyses typically assume an explicit objective, with dynamics guided by a loss minimization relative to some ground truth.  
{However, clear-cut tasks and systematic feedback are often absent in biological learning.}
Early language acquisition illustrates this regime, as infants infer statistical regularities from continuous speech before producing words and prior to explicit corrective feedback~\cite{hollich2000breaking,pinker2003language,dupoux2018cognitive,zaadnoordijk2020next}, revealing that information extraction can arise without explicit objectives.  
This motivates our central question: {\textit{Can learning spontaneously emerge and sustain itself through data exposure alone?}}

In this work, we consider a minimal model of an agent {\footnote{Throughout, ``agent'' is used in the sense of agent-based modeling~\cite{castellano2009statistical}, namely an autonomous unit governed by a local update rule, rather than in the reinforcement-learning sense of an entity that acts upon its environment.}} able to produce a classification of inputs received from a data-generating environment.
The agent's dynamics is governed by a self-consistency principle: it trains on its own predictions, while a parsimony prior penalizes complex solutions. 
As the environment continually provides new data, the agent’s internal parameters cannot strictly converge to a fixed point. 
Nonetheless, in some regimes, finite temporal correlations spontaneously emerge, indicating that the agent is able to achieve and \emph{sustain} consistent labeling. 
We refer to such steady states as \emph{functional replicators}, since the agent’s labeling scheme perpetuates itself across continual turnover. 
In simple environments, such replicators surprisingly succeed at extracting latent structure from the data as a byproduct of their stability. 

We then leverage this setting to study decentralized collective learning. {Indeed, another fundamental difference between machine and biological learning is the inherently social nature of the latter \cite{galef2005social,whiten2017social}. While some statistical mechanics studies of collective learning have started to emerge, see e.g. \citet{catania2024copycat}, these rely on interactions at the level of the \emph{weights}, which appear somewhat implausible in nature. Functional replicators, being sustained through \emph{labeling} only, offer an alternative route to the emergence of collective effects in populations of interactive learners, which we study theoretically in our simplified settings by considering agents pooling their labels given the same data samples. Crucially, this allows us to show that purely label-mediated interactions can enable spontaneous learning in regimes inaccessible to any isolated agent.}

{Our study thus contributes on two fronts. First, it shows that aligned, information-bearing behavior can ignite spontaneously from random initialization through a genuine learning phase transition, requiring neither rewards nor a selection mechanism, nor the large and diverse initial population assumed in previous work on information-processing replicators \cite{fontana1990algorithmic,y2024computational}. Second, it establishes a tractable statistical mechanics of decentralized learning in which agents coordinate through their labels alone, revealing collective regimes absent from the isolated problem. Together these connect learning theory, evolutionary principles, and social dynamics.}

\begin{figure}
\centering
\scalebox{0.75}{
\begin{tikzpicture}[font=\normalsize, every node/.style={align=center}]

\def\w{5.5}
\def\h{3.5}

\draw (0, 0) rectangle (\w, \h);
\node at (0.5*\w, \h + 0.4) {\textbf{1. Random Init ($t=0$ only)}};
\node at (1.4, 3.) {Environment $\mathcal{E}$};
\draw (0.4, 0.6) rectangle (2.4, 2.6);
\draw (4.2, 1.6) ellipse (0.8 and 0.8);
\draw[line width=1.5pt] (3.4, 1.4) -- (5.0, 1.8);
\draw[line width=1.5pt, draw=Mahogany, fill=none] (0.9, 1.6) ellipse (0.4 and 0.4);
\draw[line width=1.5pt, draw=RoyalBlue, fill=none] (1.9, 1.6) ellipse (0.4 and 0.4);
\node at (4.2, 3.) {Agent $\mathcal{W}_0$};
\path[fill=BurntOrange, opacity=0.4]
  (4.2, 1.6) -- ++(195:0.8)
  arc[start angle=195, end angle=375, x radius=0.8, y radius=0.8]
  -- cycle;

\path[fill=Emerald, opacity=0.4]
  (4.2, 1.6) -- ++(15:0.8)
  arc[start angle=15, end angle=195, x radius=0.8, y radius=0.8]
  -- cycle;

\draw[line width=1.5pt] (3.4, 1.4) -- (5.0, 1.8);

\begin{scope}[xshift=5.5cm]
\draw (0, 0) rectangle (\w, \h);
\node at (0.5*\w, \h + 0.4) {\textbf{2. Sample Mini Batch}};
\node at (1.4, 3.) {Environment $\mathcal{E}$};
\draw (0.4, 0.6) rectangle (2.4, 2.6);
\draw (4.2, 1.6) ellipse (0.8 and 0.8);
\draw[line width=1.5pt] (3.4, 1.4) -- (5.0, 1.8);
\draw[draw=gray, fill=none] (0.9, 1.6) ellipse (0.4 and 0.4);
\draw[draw=gray, fill=none] (1.9, 1.6) ellipse (0.4 and 0.4);
\node at (4.2, 3.) {Agent $\mathcal{W}_0$};
\foreach \x/\y in {
    0.7/1.4,
    0.9/1.7,
    1.1/1.85,
    1.15/1.4,
    1.0/1.6,
    0.8/1.9,
    0.9/1.3,
    0.6/1.6
  } {
    \draw[Mahogany, thick]
      (\x-0.05,\y-0.05) -- (\x+0.05,\y+0.05);
    \draw[Mahogany, thick]
      (\x-0.05,\y+0.05) -- (\x+0.05,\y-0.05);
}
\foreach \x/\y in {
    1.75/1.3,
    1.9/1.6,
    2.1/1.7,
    2.2/1.5,
    2.0/1.4,
    1.9/1.95,
    1.6/1.6,
    1.7/1.8
  } {
    \draw[RoyalBlue, thick]
      (\x-0.05,\y-0.05) -- (\x+0.05,\y+0.05);
    \draw[RoyalBlue, thick]
      (\x-0.05,\y+0.05) -- (\x+0.05,\y-0.05);
}
\path[fill=BurntOrange, opacity=0.4]
  (4.2, 1.6) -- ++(195:0.8)
  arc[start angle=195, end angle=375, x radius=0.8, y radius=0.8]
  -- cycle;

\path[fill=Emerald, opacity=0.4]
  (4.2, 1.6) -- ++(15:0.8)
  arc[start angle=15, end angle=195, x radius=0.8, y radius=0.8]
  -- cycle;

\draw[line width=1.5pt] (3.4, 1.4) -- (5.0, 1.8);
\end{scope}

\begin{scope}[yshift=-4.5cm]
\draw (0, 0) rectangle (\w, \h);
\node at (0.5*\w, \h + 0.4) {\textbf{3. Self Labelling}};
\node at (1.4, 3.) {Environment $\mathcal{E}$};
\draw (0.4, 0.6) rectangle (2.4, 2.6);
\draw (4.2, 1.6) ellipse (0.8 and 0.8);
\draw[line width=1.5pt] (3.4, 1.4) -- (5.0, 1.8);
\draw[line width=0.5pt] (0.4, 1.35) -- (2.4, 1.85);
\draw[draw=gray, fill=none] (0.9, 1.6) ellipse (0.4 and 0.4);
\draw[draw=gray, fill=none] (1.9, 1.6) ellipse (0.4 and 0.4);
\node at (4.2, 3.) {Agent $\mathcal{W}_0$};
\foreach \x/\y in {
    0.9/1.7,
    1.1/1.85,
    1.0/1.6,
    0.8/1.9,
    0.6/1.6,
    1.9/1.95,
    1.7/1.8
  } {
    \draw[Emerald, thick]
      (\x-0.05,\y-0.05) -- (\x+0.05,\y+0.05);
    \draw[Emerald, thick]
      (\x-0.05,\y+0.05) -- (\x+0.05,\y-0.05);
}
\foreach \x/\y in {
    0.9/1.3,
    0.7/1.4,    
    1.15/1.4,
    1.75/1.3,
    1.9/1.6,
    2.1/1.7,
    2.2/1.5,
    2.0/1.4,
    1.6/1.6
  } {
    \draw[BurntOrange, thick]
      (\x-0.05,\y-0.05) -- (\x+0.05,\y+0.05);
    \draw[BurntOrange, thick]
      (\x-0.05,\y+0.05) -- (\x+0.05,\y-0.05);
}
\path[fill=BurntOrange, opacity=0.4]
  (4.2, 1.6) -- ++(195:0.8)
  arc[start angle=195, end angle=375, x radius=0.8, y radius=0.8]
  -- cycle;

\path[fill=Emerald, opacity=0.4]
  (4.2, 1.6) -- ++(15:0.8)
  arc[start angle=15, end angle=195, x radius=0.8, y radius=0.8]
  -- cycle;

\draw[line width=1.5pt] (3.4, 1.4) -- (5.0, 1.8);
\end{scope}

\begin{scope}[xshift=5.5cm, yshift=-4.5cm]
\draw (0, 0) rectangle (\w, \h);
\node at (0.5*\w, \h + 0.4) {\textbf{4. Update Weights}};
\node at (1.4, 3.) {Environment $\mathcal{E}$};
\draw (0.4, 0.6) rectangle (2.4, 2.6);
\draw (4.2, 1.6) ellipse (0.8 and 0.8);
\draw[draw=gray, fill=none] (0.9, 1.6) ellipse (0.4 and 0.4);
\draw[draw=gray, fill=none] (1.9, 1.6) ellipse (0.4 and 0.4);
\node at (4.2, 3.) {Agent $\mathcal{W}_1$};
\foreach \x/\y in {
    0.9/1.7,
    1.1/1.85,
    1.0/1.6,
    0.8/1.9,
    0.6/1.6,
    0.9/1.3,
    0.7/1.4,    
    1.15/1.4
  } {
    \draw[Emerald, thick]
      (\x-0.05,\y-0.05) -- (\x+0.05,\y+0.05);
    \draw[Emerald, thick]
      (\x-0.05,\y+0.05) -- (\x+0.05,\y-0.05);
}
\foreach \x/\y in {
    1.9/1.95,
    1.7/1.8,
    1.75/1.3,
    1.9/1.6,
    2.1/1.7,
    2.2/1.5,
    2.0/1.4,
    1.6/1.6
  } {
    \draw[BurntOrange, thick]
      (\x-0.05,\y-0.05) -- (\x+0.05,\y+0.05);
    \draw[BurntOrange, thick]
      (\x-0.05,\y+0.05) -- (\x+0.05,\y-0.05);
}

\draw[->, thick, gray!70!black]
  (4.2, 1.6) ++(10:1.0) arc[start angle=10, end angle=50, radius=1.0];
\draw[->, thick, gray!70!black]
  (4.2, 1.6) ++(190:1.0) arc[start angle=190, end angle=230, radius=1.0];

\path[fill=BurntOrange, opacity=0.4]
  (4.2, 1.6) -- ++(225:0.8)
  arc[start angle=225, end angle=405, x radius=0.8, y radius=0.8]
  -- cycle;

\path[fill=Emerald, opacity=0.4]
  (4.2, 1.6) -- ++(45:0.8)
  arc[start angle=45, end angle=225, x radius=0.8, y radius=0.8]
  -- cycle;
  
\draw[line width=1.5pt, draw = gray] (3.4, 1.4) -- (5.0, 1.8);
\draw[dashed,line width=0.5pt, draw = gray] (0.4, 1.35) -- (2.4, 1.85);
\draw[line width=1.5pt, draw = black] (3.65, 1.025) -- (4.775, 2.1752);
\draw[line width=0.5pt, draw = black] ($ (0.85,1.025) - 0.38*(1.125,1.1502) $) -- ($ (1.975,2.1752) + 0.38*(1.125,1.1502) $);
\end{scope}
\end{tikzpicture}
}
\caption{Learning loop: the agent (linear in this example) assigns labels to fresh data and uses those predictions as training targets.}
\label{fig.fig0}
\end{figure}

\paragraph*{Conceptual framework.---}
We formalize our model as a system with two components:  
(i)~The \emph{environment}, a structured data generator of pairs $(\bm{x}, y)$, where $\bm{x}$ denotes the high-dimensional sample and $y$ its hidden ground-truth label.
(ii)~The \emph{agent}, a classifier with weights $\mathcal{W}_t$ and associated labelling function $\hat{y}(\bm{x};\mathcal{W}_t)$. 
The agent is randomly initialized and evolves according to a simple turnover dynamics (sketched in Fig.~\ref{fig.fig0}). At each time-step \( t \), it observes a batch of $P$ fresh samples $\{\bm{x}_{t}^\nu\}_{\nu = 1,\dots, P}$ from the environment, assigns a provisional label $\hat{y}_t^\nu$ to each, and updates its parameters by minimizing a self-consistency loss subject to a regularization on the magnitude of its parameters. 

{The resulting dynamics is governed by a competition between disordering and ordering effects. 
The continual influx of fresh samples injects noise into the agent's self-imitation loop, potentially disrupting temporal consistency and driving arbitrary fluctuations in the labeling rule. 
By contrast, weak correlations with the latent structure of the environment can be amplified in the presence of a parsimony constraint on the classifier. 
This constraint can be viewed as a minimal proxy for an energy budget on the agent's internal state, but it also plays an active role in the dynamics. Indeed, by ruling out trivial forms of self-agreement 
and favoring larger-margin classification rules, it biases the agent toward simple labeling schemes that align with the ground-truth structure of the data. 
The persistence of functional replicators across repeated turnover cycles suggests a possible learning-theoretic analogue of Dynamic Kinetic Stability~\cite{pascal2015stability,pross2022dynamic}, which has been proposed to characterize far-from-equilibrium steady states, where stability is ensured through self-replication rather than thermodynamic equilibrium.}

{Although our setup superficially resembles pseudolabeling~\cite{lee2013pseudo, arazo2020pseudo}, self-distillation~\cite{zhang2019your, takanami2025effect}, or self-training in semi-supervised learning~\cite{oymak2021theoretical}, it differs in being fully unsupervised and ground-truth-free, with each update from the very start of training relying solely on the model's own predictions on fresh data.}

\paragraph*{A concrete example.---} 
To test these ideas, we construct a minimal yet nontrivial environment in the form of a Gaussian mixture model with two symmetric components,
\begin{equation}
P(\bm{x}) = \frac{1}{2}\mathcal{N}(\bm{\mu}, \sigma^2 \mathbf{I}_N) + \frac{1}{2}\mathcal{N}(-\bm{\mu}, \sigma^2 \mathbf{I}_N),
\label{disto}
\end{equation}
where \( \bm{\mu} \in \mathbb{R}^N \) is a unit vector, and the ground-truth label \( y^\star = \pm1 \) is determined by the sign of the centroid around which \( \bm{x} \) is drawn.

For simplicity, we consider the agent to be a linear classifier, assigning binary labels using a vector of weights $\mathcal{W}_t = \bm{w}_t$ as
\begin{equation}
    \hat{y}_{t}^\nu = \mathrm{sign}(\bm{w}_t \cdot \bm{x}_{t}^\nu ), \quad \nu = 1,\dots,P
\end{equation}
to the fresh data $\bm{x}_{t}^\nu \sim P(\bm{x})$ generated at each timestep $t$.

New weights $\bm{w}_{t+1}$ are obtained via minimization of the cross-entropy loss
\begin{equation}
\label{eq:crossentropy}
\begin{aligned}
\mathcal{L}_{t}(\bm{w}_{t+1},\{\bm{x}_t^\nu, \hat{y}_t^\nu\};\lambda) = &- \sum_{\nu=1}^{P} \log h \left( \hat{y}_{t}^\nu \, \frac{\bm{w}_{t+1} \cdot \bm{x}_{t}^\nu}{\sqrt{N}} \right) \\
&+ \frac{\lambda}{2} \| \bm{w}_{t+1} \|^2,
\end{aligned}
\end{equation}
with $h(z) = 1/(1 + \e^{-z})$ the logistic function, and where $\lambda$ is the strength of the regularization term {\footnote{While the cross-entropy loss is a natural choice in this classification setting, we have also numerically experimented with other losses and have found the phenomenology to be unchanged, see Supplemental Material. Note also that our replica computation is agnostic to the choice of the loss as long as it remains convex in the learned weights.}}. 
We emphasize that, unlike standard supervised settings, the target labels in Eq.~\eqref{eq:crossentropy} are the provisional labels produced by the agent itself before the weights update, and the loss merely enforces self-consistency. 
{Note that unsupervised recovery of the signal direction in this data model was studied by \citet{biehl1994statistical}, though as a one-shot, static optimization (selecting a single weight vector that maximizes the variance or margin of the pre-activations $\bm{w}\cdot \bm{x}^\nu/\sqrt{N}$); here, by contrast, alignment is not estimated once but emerges and persists as the attractor of an iterated self-labeling dynamics.}

\paragraph*{Theoretical analysis.---}

The chosen setting allows for an exact statistical mechanics analysis of the turnover step in the thermodynamic limit, within the teacher-student framework \cite{engel2001statistical}.  
We consider the proportional regime where \( N \to \infty \), \( P \to \infty \), with \( \alpha = P/N \) fixed.  
Since we assume fresh data at each timestep, we conveniently do not have to deal with inter-step cross correlations in the dataset, as required in \cite{takahashi2022role,takanami2025effect} in the context of self-distillation \footnote{In principle, one could adapt the toolbox of \citet{takahashi2022role} to our setting to consider variations of our dynamics, the main difference being the bootstrapping of the learning process at short times.
We leave this for future work.}. 
Each step of the iteration may instead be described by computing the free entropy of the updated weights, averaged over the randomness of the inputs $\{\bm{x}_t^\nu\}$ as well as the distribution of the current configuration $\bm{w}_t$ fixing the labels; see \cite{SaraoMannelli2024Bias} and the Supplemental Material (SM) \footnote{See Supplemental Material at [URL] for further details.} for technical details.

Under the Replica Symmetric (RS) ansatz, justified here by the convex nature of the inter-step optimization problem defined by Eq.~\eqref{eq:crossentropy}, {the free entropy is given by
\begin{equation}
\label{eq:freeentropy}
\begin{aligned}
    \Phi_{t+1} &= \lim_{\beta \to \infty} \frac{1}{\beta N} \left\langle \log \int \dd \bm{w}_{t+1} \,\e^{-\beta \mathcal{L}_t(\bm{w}_{t+1},\bm{w}_t,\{\bm{x}_t^\nu\};\lambda)} \right\rangle\\
    &\sim \underset{m_{t+1},q_{t+1},R_{t+1}}{\mathrm{extr.}} s( m_{t+1},q_{t+1},R_{t};\phi_t,\alpha,\lambda,\sigma),
\end{aligned}
\end{equation}
with the order parameters $m_{t+1}$, $q_{t+1}$ and $R_{t}$ corresponding to the overlap between the updated weights and the ground truth $\bm{\mu}$, their squared norm and the overlap with the previous timestep weights respectively. }
Since the agent works with hard labels, its norm is irrelevant for the turnover trajectory. 
As a result, the labeling is fully determined by
\begin{equation}
\phi_t = \frac{\bm{w}_t \cdot \bm{\mu}}{\norm{\bm{w}_t}} = \frac{m_t}{\sqrt{q_t}},
\end{equation}
which measures the alignment of the agent to the ground truth centroid vector. We provide the full set of saddle point equations in the End Matter.

Solving the variational problem given by Eq.~\eqref{eq:freeentropy}, the agent's learning can be studied through an iteration of the form
\begin{equation}
\label{eq:m_map}
    \phi_{t+1} = f(\phi_t; \alpha, \sigma, \lambda),
\end{equation}
where \(f\) encodes the minimization of the free energy of the underlying statistical-mechanical problem {\footnote{A formally similar iterative refinement of a linear classifier on Gaussian-mixture data (albeit updated with a simpler averaging estimator) and expressed as a discrete map for the correlation with the signal direction was analyzed by~\citet{oymak2021theoretical} in a semi-supervised context, i.e. starting with a classifier pre-trained on labeled data. As such, their analysis concerns the refinement (or degradation) of its correlation to $\bm{\mu}$; the onset of learning from a vanishing correlation---the instability of the trivial fixed point and the associated learning phase transition that we study here---does not arise.}}. 
By studying the fixed points of this map, we can describe the steady states of this dynamical system and detect the emergence of functional replicators. 
{We emphasize that, although each individual turnover step is an equilibrium optimization problem, the iteration as a whole cannot be cast as the minimization of a single objective. As each step minimizes a loss built from the agent's own current labels, the dynamics is inherently out of equilibrium and trajectories need not correspond to the minimization of any global potential \footnote{A similar absence of a single global objective is well known for Kohonen's self-organizing map~\cite{kohonen1982self}, which bears some conceptual resemblance to the present setup. Indeed, \citet{erwin1992self} showed that self-organizing map updates cannot in general be written as gradient descent on any single energy function. Our turnover shares this feature, with the agent's self-generated labels acting as a moving target.}}.

\begin{figure}
\centering
\includegraphics[width=\linewidth]{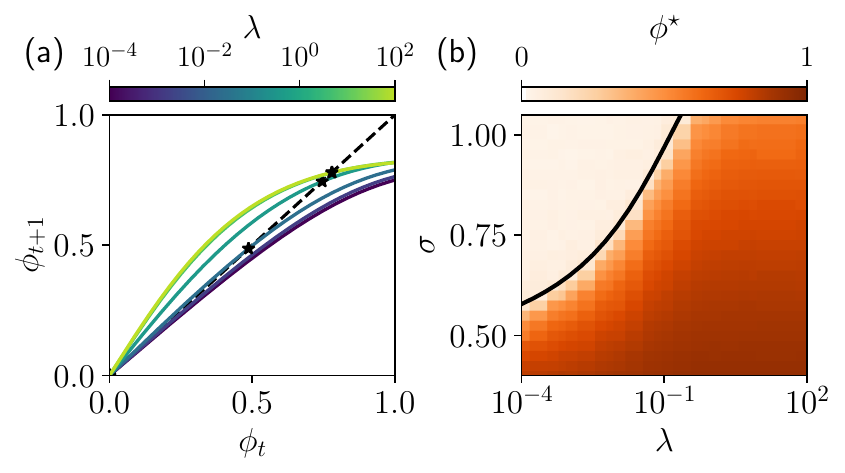}
\caption{(a) Evolution of the agent's alignment with the environment signal $\phi_t \mapsto \phi_{t+1}$ yielded by the replica analysis for $\alpha = 1$, $\sigma = 0.75$.
Star symbols indicate the fixed point solution the iterative map will reach for sufficiently large $t$.
(b) Fixed point alignment reached by numerical simulations run with $\alpha = 1$, $N = 10^3$, and averaged over 50 initial conditions, with varying values of the parameters $\lambda$ and $\sigma$.
The continuous black line shows the analytical prediction for the phase transition in the thermodynamic limit. 
}
\label{fig:master_fig}
\end{figure}

As shown in Fig.~\ref{fig:master_fig}(a), two scenarios emerge.  
When the regularization is too weak relative to the input noise, the only fixed point satisfying \(\phi^\star = f(\phi^\star)\) corresponds to \(\phi^\star = 0\).  
In other words, the agent is unable to bootstrap the learning process and extract structure from the environment, even given an infinite number of iterations.  
Conversely, when \(\lambda\) is sufficiently large, the iteration admits a nontrivial fixed point \(\phi^\star > 0\), indicating that the agent has achieved a decision rule significantly correlated with the latent structure of the environment. 
A completely equivalent and equiprobable fixed point with negative alignment (not shown in Fig.~\ref{fig:master_fig}(a)) can alternatively be reached, the sign symmetry being broken by initialization of the agent's parameters. 
Note that the only persistent channel of memory across iterations is the projection onto the signal direction, so that the long-time correlation between generations equals  \((\phi^\star)^2\) (see SM). The functional replicators that emerge in this simplified setting are thus necessarily aligned with the environment.

We analytically characterize the learning phase transition, by examining the local stability of the trivial fixed point \( \phi^\star = 0 \), governed by
\begin{equation}
r = \lim_{\phi_{t} \to 0} \frac{f(\phi_{t})}{\phi_{t}} 
\end{equation}
which can be computed from the $\phi_t = 0$ saddle point through the evaluation of two susceptibilities associated to $m_{t+1}$ (End Matter). When \( r > 1 \), small initial alignments are amplified, rendering the trivial fixed point unstable and leading to a nontrivial alignment \( |\phi^\star| > 0 \).  
The transition line \(r=1 \) in the \((\sigma, \lambda)\) plane yields a theoretical threshold that closely matches the empirical phase boundary observed in simulations, see Fig.~\ref{fig:master_fig}(b). For $\lambda \to \infty$ the criterion becomes explicit, $r^\infty = \sigma^{-2}(1+\sigma^{2})\left(1 + \pi/2\alpha\right)^{-1/2}$, so the boundary saturates at a finite $\alpha$-dependent $\sigma_c^\infty$ beyond which no regularization strength can ignite learning, see Fig.~\ref{fig:collective_static}(a) below.

Remarkably, in most signal-to-noise regimes, the steady-state alignment of the unsupervised turnover is comparable to that of supervised training with $\alpha N$ ground-truth labels, and the End Matter extends this observation to a multi-centroid, multi-label case. The SM shows further that the turnover can even surpass the supervised baseline at high noise and small $\alpha$, and that the emergence of functional replicators survives reuse of a bounded dataset, the decisive factor being the statistical richness of each update rather than a continual supply of fresh data.

\paragraph*{Collective learning in interacting agents.---}

{
As emphasized in the introduction, learning in nature is fundamentally social, spanning animal and human cognition~\cite{zaadnoordijk2020next,galef2005social,whiten2017social} and artificial swarms~\cite{long2018towards,bredeche2022social,ben2023morphological}. Since the single-agent loop closes through the agent's \textit{labels} rather than its weights, it admits a natural collective extension of self-consistent learners interacting only by exchanging the labels they assign.

We consider a population of $M$ classifying agents. At every time step, each agent is provided with an identical batch of $P$ data samples generated from the Gaussian mixture described above. The labeling for these samples, however, is different for each classifier, as the provisional labels are now pooled from a randomly assigned subset of $T$ different classifiers. 
The agents then update their weights based on the minimization of the regularized consistency objective given in Eq.~\eqref{eq:crossentropy} with the pooled labels. 

In order to study this collective setting, we adapt the replica computation described above, with the order parameter updates of agent $i$ depending on the alignments of all $T$ randomly selected teachers. 
Computing the associated free entropy thus also involves all the overlaps between the updated weights of $i$ and those of its teachers. 
To identify the onset of learning, as well as the collective fixed point alignment, it suffices to consider a scenario where all agents have a homogeneous alignment with the signal---$\phi_t^i = \phi_t$ $\forall i$---and independent orthogonal components. 
As a result, the saddle-point equations underpinning the long-time average alignment in the population closely resemble those of the single-agent setting, with an analytic dependency on $T$. 
These equations also admit a clean $T\to\infty$ limit upon proper rescaling of the teacher-student overlap. 
Under this homogeneity assumption, the alignment thus follows the iteration 
\begin{equation}
    \phi_{t+1} = f_T(\phi_t;\alpha,\sigma,\lambda),
\end{equation}
where $f_1$ recovers the single-agent evolution of the previous section (details in the End Matter).

As in the single-agent case, the spontaneous learning of the data model can be studied through the fixed points of this map. 
We first show the modification of the phase boundaries, identified through the slope of the $f_T$ at the origin that can again be computed from the $\phi_t = 0$ saddle, in Fig.~\ref{fig:collective_static}(a). 
At large values of $\sigma$, where learning the underlying structure is challenging due to a small signal to noise ratio, pooling labels across different teacher networks significantly enlarges the region where spontaneous learning is possible. This effect can be made explicit in the $\lambda\to \infty$ limit of the slope, leading to a $T$-dependent ceiling on the data variance (SM), 
\begin{equation}
\left(\sigma_c^\infty(T)\right)^2 = \bigg(\sqrt{\frac{\pi}{2\alpha} + \frac{1}{T}} - 1\bigg)^{-1},
\end{equation}
shown for $\alpha = 1$ by the horizontal dashed lines in Fig.~\ref{fig:collective_static}(a). Remarkably, this expression shows that if $\alpha > \pi/2$, there exists a pool size $T^\star = 2\alpha/(2\alpha - \pi)$ above which learning can always be achieved for sufficiently large $\lambda$.

In addition to moving the phase boundary, we show in Fig.~\ref{fig:collective_static}(b) that collective pooling of labels may also significantly increase the fixed-point value of the alignment to the optimal solution. Fig.~\ref{fig:collective_static}(b) also shows the very good agreement of our homogeneous theory's prediction of the fixed point alignment with finite size numerical simulations for $T>1$ (see below for the specificity of the $T=1$ case). 
The positive effect of pooling in this region can finally be seen by inspecting the map itself, as shown in Fig.~\ref{fig:collective_static}(c) right-most panel, illustrating a case where a non-zero fixed point only exists for $T > 1$, demonstrating that interaction can genuinely enable learning.

Pooling is not always beneficial, however. 
On the contrary, when the signal to noise ratio is high (small $\sigma$), we observe a significant increase in the value of regularization required to observe spontaneous alignment to the optimal solution. 
Intuitively, the phenomenon may be understood as follows. 
Close to $\phi_t = 0$, all $T$ teachers will have a very small overlap to each other of $O(\phi_t^2)$. 
As a result, the labeling inconsistencies introduced by the randomly initialized teachers make each pooled datasets effectively non-linearly separable for the students. 
In this regime, only sufficiently strong regularization can suppress the conflicting directions induced by the different teachers and allow the iterative dynamics to converge toward the aligned solution. 
This phenomenon interestingly leads to the appearance of an unstable fixed point in the iterative map when both $\sigma$ and $\lambda$ are small, and the single-agent iteration manages to converge to a large alignment at the fixed point. 
As illustrated in Fig.~\ref{fig:collective_static}(c), left-most panel, while the slope of the map is insufficient to initiate learning close to $\phi_t = 0$, a warm-start initialization may still converge to $\phi^\star > 0$, reaching a smaller but comparable alignment to the single-agent fixed point. 
As the problem becomes more difficult and a single agent converges to a smaller $\phi^\star$, the collective learning system undergoes a saddle-node bifurcation and the unstable fixed point disappears, meaning that the pool of teachers is already too large to sustain a \emph{functional replicator} state, regardless of the initialization. 

\begin{figure}
    \centering
    \includegraphics[width=\linewidth]{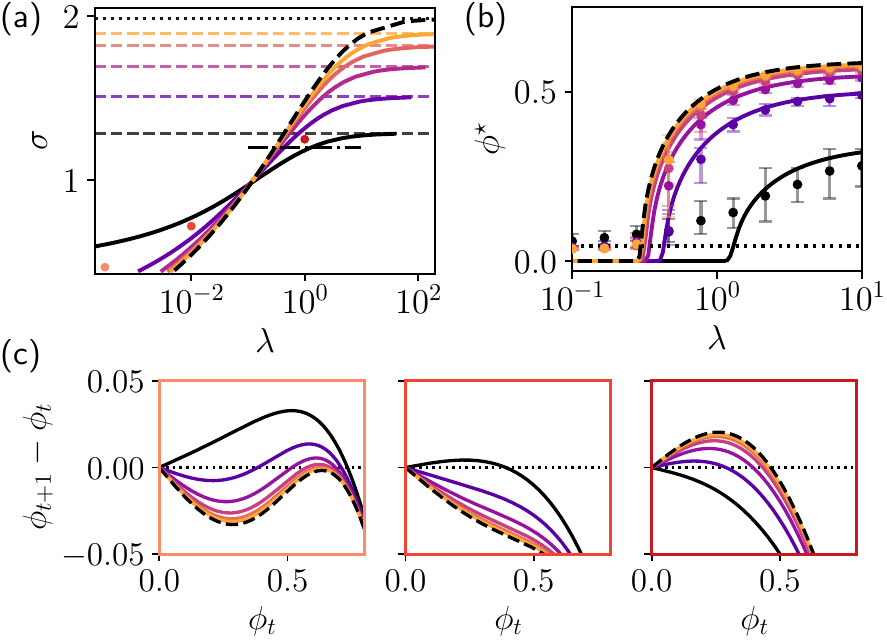}
    \caption{{Replica theory of collective learning with label pooling from $T = \{1,2,4,8,16,32\}$ teachers (from black to light orange), dashed lines representing the $T\to \infty$ limit. (a) Phase boundary in the $(\sigma,\lambda)$ plane ($\alpha = 1$) separating $\phi^\star = 0$ from $\phi^\star > 0$. (b) Fixed-point alignment $\phi^\star$ along the slice indicated by the dashed-dotted line in (a). Solid lines: replica theory under the homogeneous ansatz. Markers: numerical simulations with $N = P = 500$, $M = 64$ agents, simulated for 400 population sweeps and averaged over 16 realizations. Dotted line: $1/\sqrt{N}$ showing the scale of the expected finite size remanent alignment around the $\phi^\star = 0$ fixed point. (c) Increment $\phi_{t+1} - \phi_t$ versus $\phi_t$ for the three parameter pairs marked in by the colored markers in (a). Horizontal dotted lines mark a fixed point $\phi_{t+1} = \phi_t = \phi^\star$.}}
    \label{fig:collective_static}
\end{figure}

\begin{figure}
    \centering
    \includegraphics[width=\linewidth]{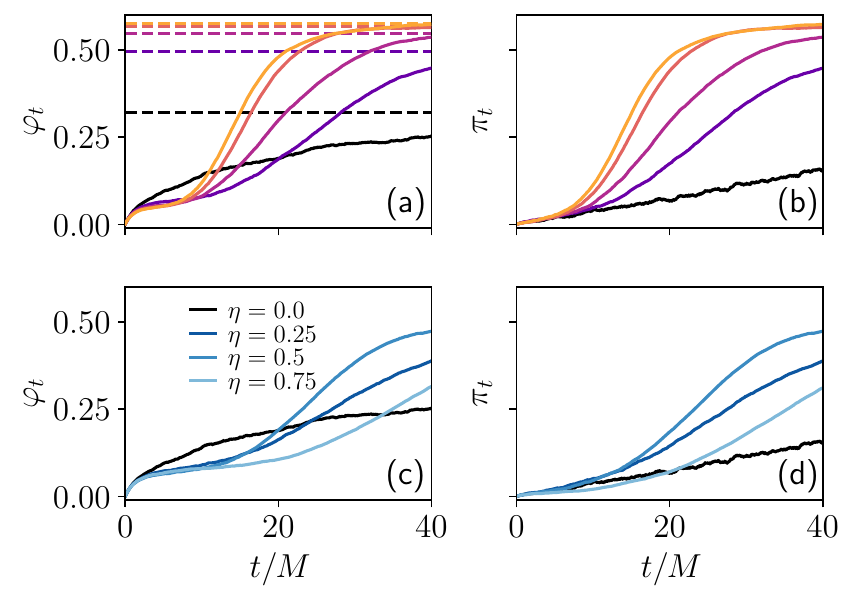}
    \caption{Agent-based modeling of the collective learning setting for $N = P = 500$ ($\alpha = 1)$, $\sigma = 1.2$, $\lambda = 10$ and $M = 64$ agents. Top: evolution of (a) the average absolute alignment, dashed line showing the replica theory fixed-point prediction, and (b) consensus level for $T = \{1,2,4,8,16,32\}$ (from black to light orange). Bottom: (c) and (d) identical to (a) and (b) respectively but fixing $T=1$ and varying the ``stubbornness'' of the agent $\eta$, representing the fraction of its own label it mixes with that of the teacher, from 0 (as in (a) and (b)) to 0.75.
    }
    \label{fig:ABM_experiments}
\end{figure}

While the homogeneity assumption $\phi_t^i = \phi_t$ allowed us to correctly identify the onset of collective spontaneous learning as well as the fixed point alignment $\phi^\star$, it cannot capture the \textit{dynamics} from a randomly initialized population to functional replicators. 
As mentioned above, the learning problem admits a sign symmetry $\bm{w}_t \mapsto-\bm{w}_t$. 
Although this symmetry was inconsequential for the evolution of $\phi_t$ in the single-agent case, it plays a crucial role in the transient dynamics of collective systems. }
We introduce two key observables: the average \textit{individual alignment}, defined as  
$\varphi_t = \frac{1}{M} \sum_i |\phi_t^i|$, and the \textit{consensus level}, defined as  
$\pi_t = \left| \frac{1}{M} \sum_i \phi_t^i \right|$. 
As such, having $\pi_t \to \varphi_t$ signals the convergence to a homogeneous state $\phi_t^i = \phi_t$.

We illustrate the evolution of these two quantities obtained from numerical simulations for different values of $T$ in Fig.~\ref{fig:ABM_experiments}, top row. 
In addition to the difference in steady-state alignment, it is clear that the dynamics are qualitatively distinct as the pool size is varied. 
The case of $T=1$, in particular, displays an extremely slow convergence to consensus, and therefore to the homogeneous fixed point, compared to its $T>1$ counterparts, explaining the recorded discrepancy between simulations and theoretical predictions in Fig.~\ref{fig:collective_static}(c), as the numerics fail to converge despite the large iteration cutoff. 
These slow dynamics can be explained by exploiting results from the statistical mechanics of consensus dynamics \cite{castellano2009statistical}. 
Considering the mapping to a spin system with variables $\sigma_i^t = \sign(\phi_i ^t)$, the model can be interpreted as a variation of a mean-field voter model, or more precisely to the Moran model from population genetics~\cite{moran1958random,blythe2007stochastic}. 
As such, the dynamics of $\pi_t$ are purely noise-driven (as visible from the rugged nature of the curves), and consensus is reached on a slow $O(M^2)$ timescale. 
Increasing the number of teachers in the pool induces a drive towards the mean alignment, accelerating the symmetry breaking and allowing for a much faster $O(M)$ convergence time. 
In the mean-field limit $T\to \infty$, the teacher-sampling fluctuations should vanish and the population collapses onto the homogeneous deterministic iteration after a single step.

The picture admits a natural enrichment by reintroducing the agent's own labels into the pool, weighted by a stubbornness parameter $\eta \in [0, 1]$. At each step, the student adopts its own provisional labels for a fraction $\eta$ of the batch and pools labels from $T$ other teachers for the remaining fraction. 
The cases discussed so far correspond to $\eta = 0$, while $\eta = 1$ recovers the independent single-agent setting (rendering consensus unreachable). 
Crucially, $\eta$ provides a continuous knob to tune the balance between individual learning and collective imitation, and consequently the nature of the consensus dynamics. 
As shown in Fig.~\ref{fig:ABM_experiments}, bottom row, even modest values of $\eta < 1/2$ are sufficient to break the Moran-like fluctuation-driven regime in the hard learning task, restoring an $O(M)$ convergence to consensus. 
A description of the dynamical crossover across learning regimes is provided in the End Matter. 
Beyond this specific extension, the framework readily accommodates further variations (heterogeneous interaction networks or pool sizes, asymmetric stubbornness), illustrating the potential of self-consistent learners as a minimal building block at the interface of learning theory and collective dynamics.

\paragraph*{Conclusion.---} 
In this Letter, we have shown that minimal self-consistent agents can evolve into \emph{functional replicators}, sustaining coherent internal states and spontaneously aligning with the latent structure of a data-producing environment.   
In multi-agent settings, where agents interact solely through the labels they exchange, collective pooling {reshapes the learning phase boundary, in some regimes enabling learning inaccessible to any isolated agent and suppressing it in others, while connecting to standard consensus dynamics.}

Beyond the specific scenarios studied here, our approach opens a route to explore learning as an emergent and self-organized process, bridging ideas from statistical mechanics, evolutionary theory, and artificial intelligence.
At the single agent level, key open questions include how such replicators behave under temporally structured inputs, and how richer architectures or environments alter the landscape of attractors. 
Considering populations of learning agents, we believe that spatially extended versions of our model, with potential motile agents, could provide a promising setup for studying intelligent active matter in the spirit of the recent work of \citet{jung2025kinetic}.

\paragraph*{Acknowledgments.---}
The authors are grateful to B.~L. Annesi, C. Baldassi, O. Dauchot, T. Takahashi, and R. Zakine for fruitful discussions. L.S. and S.A. acknowledge the support of the European Union - Next Generation EU funds, CUP J53D23001390001, N. Prot.2022E3WYTY. 
The work of J.\,G.-B. was supported by the European Union’s Horizon Europe program under the Marie Skłodowska-Curie grant agreement No.~101210798.

\bibliography{bibs}

\section{End Matter}

{
\subsection{Replica theory saddle-point equations}
The variational form Eq.~\eqref{eq:freeentropy} reduces to the extremization of a free entropy density over the order parameters $\{m_{t+1}, q_{t+1}, R_{t}, \delta q_{t+1}\}$ and their conjugates $\{\hat{m}_{t+1}, \hat{q}_{t+1}, \hat{R}_t, \delta \hat{q}_{t+1}\}$. The full computation is given in the SM, we here provide the final set of equations in the $N \to \infty$, $\beta \to \infty$ limit.

Defining $\Phi_{t+1} = \operatorname{extr.} \,g_I + g_S + \alpha g_E$, the three contributions read (we suppress the $t+1$ subscript on order parameters and write $\phi_t \equiv \phi$ for the teacher alignment):
\begin{equation}
\begin{aligned}
g_I &= - \left[\hat{m}m + \hat{R} R + \frac{1}{2}(\hat{q} \delta q - \delta \hat{q} q) \right], \\
g_S &= \frac{(\hat{m} + \phi \hat{R})^2 + \hat{q} + (1-\phi^2)\hat{R}^2}{2(\lambda + \delta \hat{q})},\\
g_E &= \frac{1}{2} \sum_{c=\pm 1} \int  Dz\, \Big[ H(-A_c - B z)\, \mathcal{M}(c, z; 1) \\
&\qquad +  H(A_c + B z)\, \mathcal{M}(c, z; 0) \Big],
\end{aligned}
\end{equation}
where $Dz = \e^{-z^2/2}/\sqrt{2\pi}\,\dd z$, $H(x) = \int_x^\infty Dz$, the auxiliary fields are
\begin{equation}
A_c = \frac{c \phi\sqrt{q}}{\sigma\sqrt{q - R^2}}, \qquad B = \frac{R}{\sqrt{q-R^2}},
\end{equation}
and the inner extremization is
\begin{equation*}
\mathcal{M}(c, z; y) = \max_{u} \left[-\frac{u^2}{2} - \ell \left(y, \sigma\sqrt{\delta q}u + c m + \sigma\sqrt{q} z\right)\right],
\end{equation*}
with $\ell$ being the logistic loss here, $\ell(y,h) = \log (1+\e^{-yh})$.

The saddle-point equations are obtained by extremizing $\Phi_{t+1}$ with respect to all order parameters. Differentiating $g_S$ yields the equations for the next step order parameters,
\begin{equation}
\begin{aligned}
m &= \frac{\hat{m} + \phi \hat{R}}{\lambda + \delta \hat{q}}, \qquad
R = \frac{\phi(\hat{m}+\phi\hat{R}) + (1-\phi^2)\hat{R}}{\lambda + \delta \hat{q}}, \\[2pt]
q &= \frac{(\hat{m}+\phi\hat{R})^2 + \hat{q} + (1-\phi^2)\hat{R}^2}{(\lambda + \delta \hat{q})^2}, \quad
\delta q = \frac{1}{\lambda + \delta \hat{q}},
\end{aligned}
\end{equation}
while differentiation of $g_E$ gives the conjugate equations,
\begin{equation}
\begin{aligned}
\hat{m} &= \alpha \partial_m g_E, \qquad \hat{R} = \alpha \partial_R g_E, \\
\hat{q} &= 2\alpha \partial_{\delta q} g_E, \qquad \delta \hat{q} = -2\alpha \partial_q g_E.
\end{aligned}
\end{equation}
A Julia code solving these equations efficiently is available on request.}

The learning transition follows from linearizing the map resulting from this variational problem at $\phi = 0$. The free entropy is invariant under $(\phi, m, \hat{m}) \to (-\phi, -m, -\hat{m})$, so along the trivial branch all order parameters except $(m, \hat{m})$ are even in $\phi$, and the linear response closes on this pair. Following the computation detailed in the SM and denoting by a subscript $0$ the $\phi = 0$ saddle, we find
\begin{equation}
r = \lim_{\phi \to 0} \frac{f(\phi)}{\phi} = \frac{1}{\sqrt{q_0}} \frac{\partial m}{\partial \phi} \bigg|_{0} = \frac{\hat{R}_0 + \alpha \chi_{m\phi}}{\sqrt{q_0}(\lambda + \delta\hat{q}_0 - \alpha \chi_{mm})},
\label{eq:r_closed}
\end{equation}
with $\chi_{mm} = \partial^2_{mm} g_E|_0 \leq 0$ and $\chi_{m\phi} = \partial^2_{m\phi} g_E|_0 \geq 0$ given as one-dimensional Gaussian integrals in the SM. The two terms in the numerator identify the transmission channels of the alignment, the persistence of the previous weights and the signal bias of the self-generated labels. For $\lambda \to \infty$ one has $\chi_{mm} \to -1/4$, $\chi_{m\phi} \to 1/(\sigma\sqrt{2\pi})$, $\hat{R}_0 \to \alpha\sigma/\sqrt{2\pi}$, $\delta\hat{q}_0 \to \alpha\sigma^2/4$ and $\lambda^2 q_0 \to \alpha\sigma^2(1/4 + \alpha/2\pi)$, yielding the expression for $r^\infty$ quoted in the main text and the saturation value $(\sigma_c^\infty)^2 = (2\alpha/\pi)(1 + \sqrt{1 + \pi/2\alpha})$.

\subsection{Extension to richer data models} 

\begin{figure}
    \centering
    \includegraphics[width=0.8\linewidth]{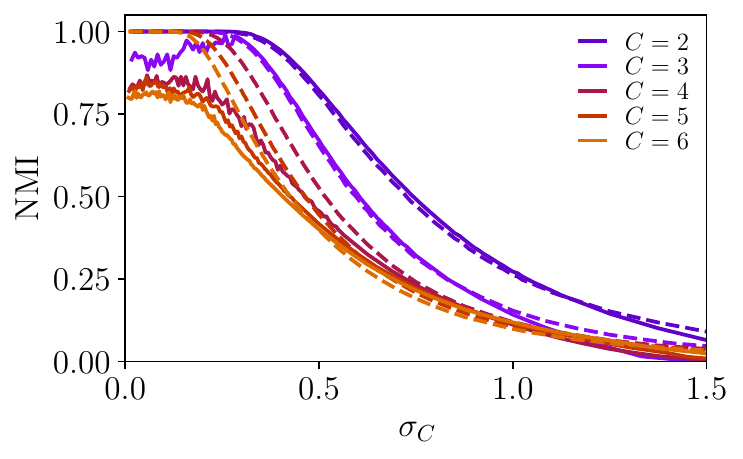}
    \caption{Steady state Normalised Mutual Information (NMI) between the learned and true labels for data distributions with $C\geq 2$ centroids of homogeneous standard deviation $\sigma_C = \sigma/C$, averaged over 50 numerical simulations, with $ \alpha_C= \alpha C$, $\alpha = 1$, $N = 10^3$ and $\lambda= 10^2$. Dashed lines show the NMI achieved by a traditional supervised training procedure with $\alpha_C  N$ labeled examples.}
    \label{fig:multicentroid}
\end{figure}

To test the robustness of our framework, we also explore a multiclass scenario with \( C > 2 \) classes.  
The environment is defined as a mixture of \( C \) Gaussian components:
\begin{equation}
P(x) = \frac{1}{C} \sum_{c=1}^C \mathcal{N}\left(\bm{\mu}_c, \sigma_C^2 \mathbf{I}_N\right),
\end{equation}
where the centroids \( \{\bm{\mu}_c\} \) are placed as equidistant points on the unit circle in \( \mathbb{R}^2 \), and then embedded into \( \mathbb{R}^N \).  
To ensure a fair comparison across different values of \( C \), we scale the homogeneous variances as \( \sigma_C^2 = \sigma^2 / C^2 \), and adjust the dataset size by setting \( P_C = P \times C \). In alternative to the alignment---promoted to a matrix in the multiclass setting---we employ the Normalised Mutual Information (NMI) between the model’s predictions \( \hat{y} \) and the true labels \( y^\star \) (a definition is provided in the SM). 
This metric ranges from 0---random labeling---to 1---perfect ground-truth rule recovery up to label permutation.

The replica computation can be extended to this setting, but the resolution of the associated saddle-point equations becomes considerably more costly~\cite{Loureiro2021GMM}. We instead explore the phenomenology through direct numerical simulation of large systems. As shown in Fig.~\ref{fig:multicentroid}, functional replicators still emerge in the multiclass setting under strong regularization. Here the NMI is in general not self-averaging over initial conditions but develops distinct modes set by the level of centroid coverage. A lower average NMI may thus reflects settings where the maximum is reached only by a few fortunate initializations, the rest converging to imperfect coverage. The corresponding variances are omitted from Fig.~\ref{fig:multicentroid} for readability, with the best outcomes per setting reported in the SM.

For intermediate $\sigma_C$, the average steady-state NMI is competitive with that of supervised training on ground-truth labels. As $C$ increases, however, the average NMI deteriorates and the range of $\sigma_C$ over which the agent recovers the full latent structure narrows, as the dynamics can be drawn to initialization-dependent spurious steady states that remain robust across iterations but employ fewer labels than the ground-truth distribution. At low to intermediate noise, this tendency can be countered by overparametrizing, i.e.\ taking an output dimension $K > C$ (SM), whereby the agent then exploits the extra dimensions to achieve full coverage, discarding the excess labels. Much as in the lottery-ticket phenomenon~\citep{frankle2018lottery}, redundancy and diverse initializations promote spontaneous adaptation.

{Beyond this idealized mixture of Gaussians, we also performed numerical experiments using samples from the MNIST dataset \cite{mnist}, see SM. There again, we find that the unsupervised turnover dynamics, when appropriately regularized, achieve good performance despite never being exposed to ground-truth labels.}


{
\subsection{Collective learning}

\paragraph{Replica theory saddle-point.---} As mentioned in the main text, the replica theory may be adapted to a pool of $T$ teachers. Under the homogeneity assumption that all teachers have identical alignment, the saddle-point structure derived above is preserved, with the only modification entering through the multiplicity of teacher contributions in the entropic term $g_S$,
\begin{equation}
g_S^{(T)} = \frac{(\hat{m} + T\phi \hat{R})^2 + \hat{q} + T(1-\phi^2)\hat{R}^2}{2(\lambda + \delta \hat{q})},
\end{equation}
while we have the straightforward substitution $R\hat{R} \to TR \hat{R}$ throughout $g_I$ and $g_E$.
Correspondingly, the equations for $m$, $R$, and $q$ acquire $T$-dependent contributions,
\begin{equation}
\begin{aligned}
m &= \frac{\hat{m} + T\phi\,\hat{R}}{\lambda + \delta \hat{q}}, \qquad
R = \frac{\phi(\hat{m}+T\phi\hat{R}) + (1-\phi^2)\hat{R}}{\lambda + \delta \hat{q}}, \\
q &= \frac{(\hat{m}+T\phi\hat{R})^2 + \hat{q} + T(1-\phi^2)\hat{R}^2}{(\lambda + \delta \hat{q})^2}.
\label{eq:saddle_OP}
\end{aligned}
\end{equation}
The conjugate equations for $\hat{m}, \hat{q}, \delta \hat{q}$ are unchanged, while $\hat{R}$ inherits a $1/T$ factor from the modified interaction term,
\begin{equation}
\hat{R} = \frac{\alpha}{T}\partial_R g_E.
\end{equation}
The iterated map $\phi_{t+1} = f_T(\phi_t;\alpha,\sigma,\lambda)$ then follows from the self-consistent solution at each step, recovering the single-agent equations of the previous section for $T = 1$. 

In the $T\to\infty$, the contributions scaling with $T$ in $g_S^{(T)}$ dominate, and introducing the natural rescaling $\hat{R}' = T\hat{R}$ produces a closed limiting system in which the teacher-student overlap effectively decouples from the student's self-overlap. 
At finite $T$, corrections to this limit scale as $1/T$, explaining the rapid saturation of the gain observed numerically in Fig.~\ref{fig:collective_static}. Specifically, the corrections can be identified as the ``variance'' terms entering Eq.~\eqref{eq:saddle_OP} in the expressions of $R$ and $q$ (prefactored by $1-\phi^2$). These contributions, which may be attributed to the the teacher-specific random component orthogonal to the ground truth, are thus naturally averaged out when the number of teachers in the pool diverges, leaving only the coherent part related to $m$.} 

The stability analysis yielding $r(T)$ extends verbatim, $g_E$ being unaffected by pooling. The criterion keeps the form of Eq.~\eqref{eq:r_closed} with $\hat{R}_0 \to T\hat{R}_0$ and the susceptibilities evaluated at the $T$-dependent trivial saddle, whose disorder contribution $T\hat{R}_0^2 \propto 1/T$ is averaged out by the pool. For $\lambda \to \infty$ this gives 
\begin{equation}
    r^\infty(T) = \frac{1+\sigma^{2}}{\sigma^2}\left(\frac{\pi}{2\alpha} + \frac1T\right)^{-1/2},
\end{equation}
giving the saturation value in the main text by setting $r_c^\infty(T) = 1$.

\begin{figure}[b]
    \centering
    \includegraphics[width=\linewidth]{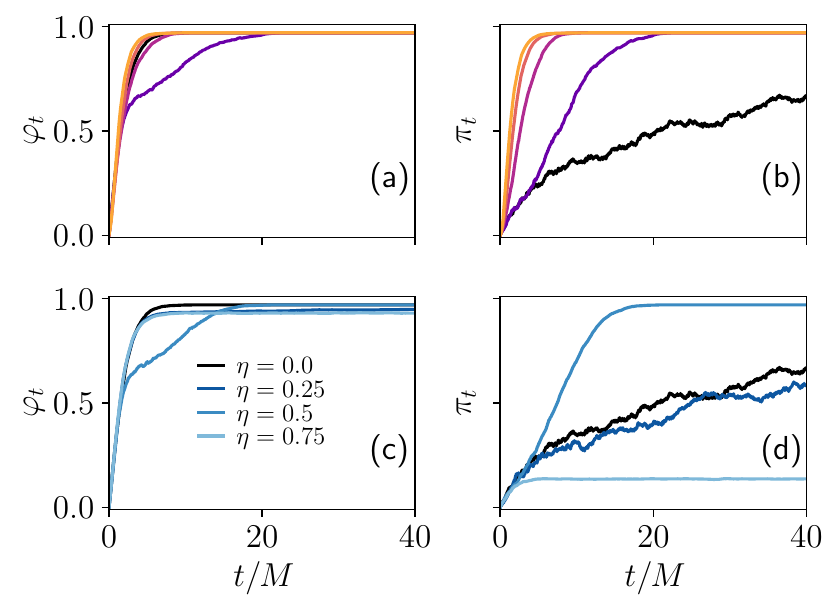}
    \caption{Agent-based modeling of the collective learning setting for $N = P = 500$ ($\alpha = 1)$, $\sigma = 0.25$, $\lambda = 10$ and $M = 64$ agents. Top: evolution of (a) the average absolute alignment and (b) consensus level for $T = \{1,2,4,8,16,32\}$ (from black to light orange). Bottom: (c) and (d) identical to (a) and (b) respectively but fixing $T=1$ and varying the ``stubbornness'' of the agent $\eta$, representing the fraction of its own label it mixes with that of the teacher, from 0 (as in (a) and (b)) to 0.75.
    }
    \label{fig:ABM_experiments_easy}
\end{figure}

\paragraph*{Consensus dynamics.---} 
The collective dynamics depend not only on the pool size $T$ and stubbornness $\eta$, but crucially on the difficulty of the underlying learning problem itself. 
For $T=1$, $\eta=0$, the sign dynamics map exactly onto a neutral Moran process~\cite{moran1958random,blythe2007stochastic}, with consensus reached on $O(M^2)$ timescales regardless of $\sigma$. 
The case $T=2$ or $T=1$ with $\eta > 0$ reveals a richer interplay. 
In the easy regime (small $\sigma$, Fig.~\ref{fig:ABM_experiments_easy}), individual agents reach $|\phi_t^i| \sim 1$ within a few iterations, well before any collective alignment has had time to form. 
A student with $T = 2$ teachers then typically faces two teachers of opposite polarization, breaks the tie at random, and the population-level dynamics reduces to a Moran-like fluctuation-driven process on the slow $O(M^2)$ timescale initially. 
Once symmetry is spontaneously broken, the majority drift takes over and consensus completes rapidly, as shown in Fig.~\ref{fig:ABM_experiments_easy}(a)-(b). The picture is qualitatively similar with $T=1$ and finite stubbornness $\eta$, although the additional parameter affords more nuance.
For $\eta \leq 1/2$, consensus eventually emerges, but the timescale depends sharply on $\eta$; the case $\eta = 1/2$ achieves the fastest convergence to consensus despite producing the slowest growth of individual alignment, while smaller values of $\eta$ recover the $O(M^2)$ Moran-like scaling. 
For $\eta > 1/2$, consensus fails to develop beyond the small residual alignment built up during the initial common learning phase, with $\pi_t$ saturating strictly below $\varphi_t$.

This separation of timescales between individual learning and collective coordination disappears in the hard regime (large $\sigma$, main text Fig.~\ref{fig:ABM_experiments}). There, learning and consensus formation evolve in tandem, and the persistent presence of an aligned majority during the build-up of $|\phi_t^i|$ is sufficient to drive consensus on the much faster $O(M)$ scale, paradoxically accelerated by the slowness of individual learning.
This trade-off between individual learning and collective coordination is reminiscent of phenomena in Schelling-type models~\cite{schelling2006micromotives,grauwin2009competition,garnier2024hydro}, and suggests that coupling learning dynamics to consensus mechanisms may produce a richer phenomenology than either ingredient in isolation.

\end{document}


\title{Supplemental Material for ``Spontaneous and Collective Learning in a Minimal Agent-Environment Model''}

\author{Sebastiano Ariosto}
\email{sebastiano.ariosto@unibocconi.it}
\author{Jérôme Garnier-Brun}
\email{jerome.garnier@unibocconi.it}
\author{Luca Saglietti}
\author{Davide Straziota}
\affiliation{%
Bocconi Institute for Data Science and Analytics (BIDSA) \& Department of Computing Sciences, Bocconi University, Milan, Italy
}

\date{\today}

\maketitle

\tableofcontents

\renewcommand{\theequation}{S\arabic{equation}}
\renewcommand{\thetable}{S\arabic{table}}
\renewcommand{\thefigure}{S\arabic{figure}}
\setcounter{equation}{0}
\setcounter{table}{0}
\setcounter{figure}{0}

\renewcommand{\thesection}{\Roman{section}}

\section{Additional results}

\subsection{Learning dynamics}

In the main text we focused on the stationary properties of the agent, showing how the long-term behavior is captured by the fixed point of the recursion in Eq.~\eqref{eq:m_map} of the main text. 
It is worth emphasizing, however, that this map is not restricted to the stationary regime, as it provides the correct update rule at each step of the dynamics, provided that each turnover update reaches the loss minimizer.

\begin{figure}
    \centering
    \includegraphics[width=0.8\linewidth]{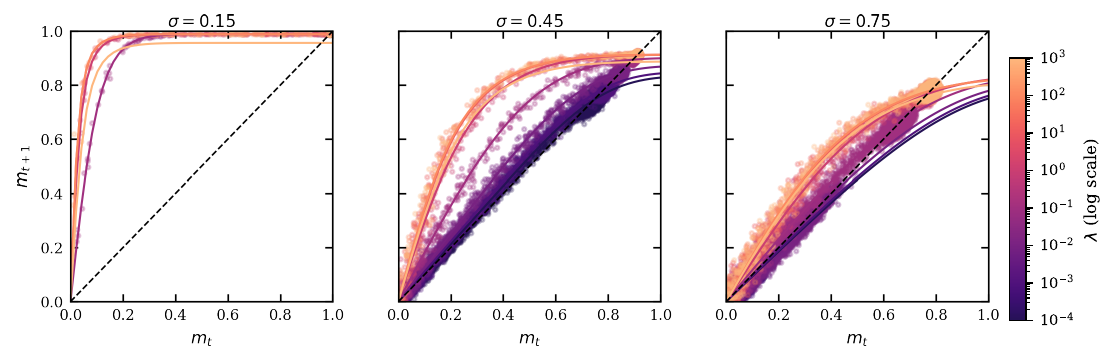}
        \caption{Learning dynamics across noise regimes, here for normalized weights $m_t = \phi_t$, $q_t = 1$.
        Comparison between the agent's inferred centroid positions \( m_t \) and the true class centroids \( \phi^\star_t \), for increasing noise levels \( \sigma = 0.15, 0.45, 0.75 \). 
        Each panel shows the learning trajectory (color-coded by \( \lambda \), on a logarithmic scale), indicating convergence toward a dynamical attractor.
        As \( \sigma \) increases, the agent's representation becomes noisier, yet structured learning persists in early principal directions.}
\label{fig:m_map}
\end{figure}

Figure~\ref{fig:m_map} confirms that this theoretical prediction accurately describes the transient dynamics across a wide range of parameters $\sigma$ and $\lambda$. 
The scattered cloud of points around the optimal $\phi$ values indicates that the system does not converge to a single deterministic configuration, but rather to a dynamical attractor characterized by persistent finite-size fluctuations around the theoretical fixed point. 
Despite these fluctuations, the dynamics retain functional stability and sustain high performance, as highlighted by the large values of $\phi$ reached.

\begin{figure}
    \centering
    \includegraphics[width=0.8\linewidth]{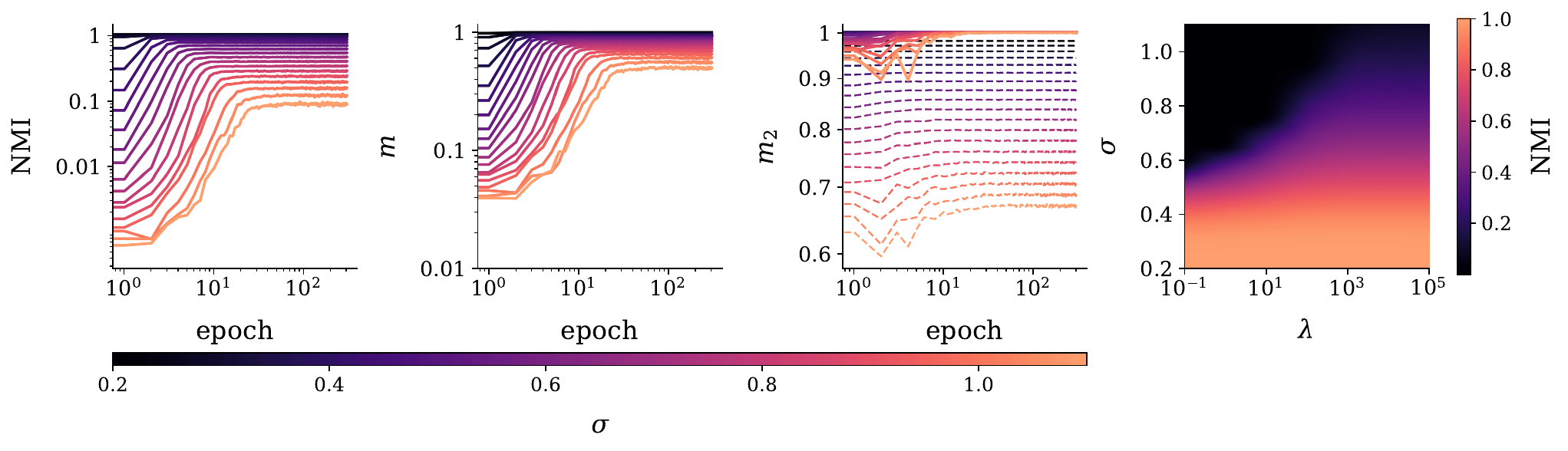}
        \caption{Evaluation of structure extraction and robustness to noise, here for normalized weights $m_t = \phi_t$, $q_t = 1$. 
        (Left to right) Normalized Mutual Information (NMI) between learned and true labels; overlap between agent weights \( \bm{w}_t \) and data centroids \( \bm{\mu} \); overlap restricted to top two principal components; NMI as a function of the noise level \(\sigma\) and the regularization strength \( \lambda \). 
        Results show that while the full overlap degrades with increasing noise, the top components remain robust, and accuracy remains high in a broad region of the \((\alpha, \lambda)\) plane.}
\label{fig:noisy_dynamics}
\end{figure}

This behavior is further illustrated in Fig.~\ref{fig:noisy_dynamics}. 
As the input noise $\sigma$ increases, the total overlap between the agent’s weights and the data centroids decreases, reflecting the growing difficulty of extracting information from noisy environments. 
Crucially, when the analysis is restricted to the first two principal directions of the data—the only ones relevant for this binary classification task—the overlap $m_2$ remains close to unity up to relatively large values of $\sigma$. 
This shows that the degradation of NMI at high noise does not stem from a failure to capture the latent structure, but rather from spurious alignments with noisy components. 
For comparison, in the supervised case (dashed lines, with $\alpha = 1$), the overlap along the two relevant directions decreases significantly faster, indicating a weaker robustness to noise. 
Note, however, that the comparison is not fully fair: in the turnover dynamics, the effective number of patterns seen grows with the number of epochs, i.e. $\alpha = \#\text{epochs}$.

\begin{figure}
    \centering
    \includegraphics[width=\linewidth]{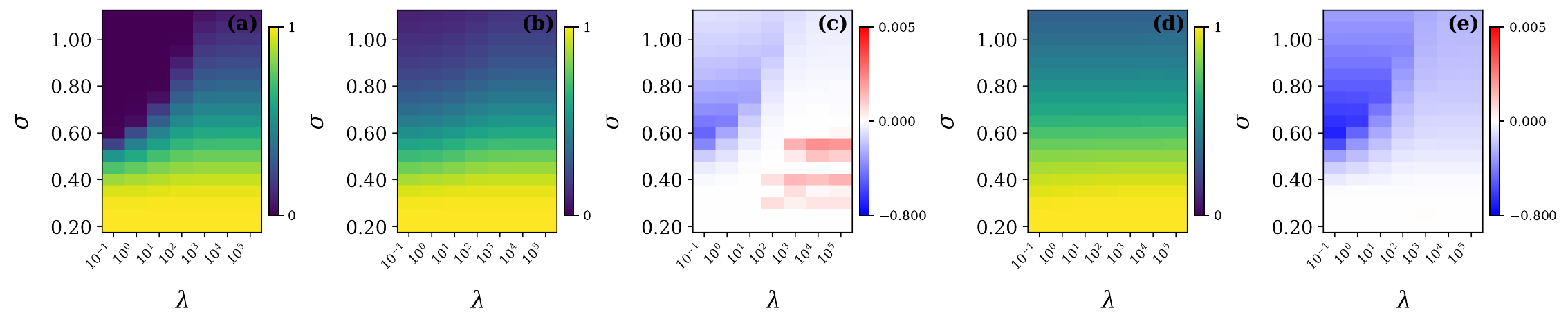}
    \caption{%
    Comparison between the normalized mutual information (NMI) obtained by the turnover dynamics and by supervised training.  
    (a) Steady-state NMI for the unsupervised turnover dynamics as a function of $\sigma$ and $\lambda$.  
    (b) NMI achieved by a supervised classifier trained on $\alpha=1$ labeled examples.  
    (c) Difference $\Delta\text{NMI} = \text{NMI}_{\text{turnover}} - \text{NMI}_{\text{supervised}}$ for $\alpha=1$, highlighting a relative advantage of the unsupervised dynamics at intermediate noise and large regularization.  
    (d) NMI of a supervised classifier with $\alpha=100$ labeled examples.  
    (e) Difference $\Delta\text{NMI}$ for $\alpha=100$, showing close agreement between the two approaches, with the turnover procedure exhibiting comparable performance in certain regions of parameter space.  
    All results are averaged over 50 independent runs with N = $10^3$; the supervised baselines use $\alpha$ = 1 in panels (b,c) and $\alpha$ = 100 in panels (d,e).}
\label{fig:NMI_super}
\end{figure}

A direct benchmark is provided in Fig.~\ref{fig:NMI_super}, where we contrast the performance from the turnover dynamics with that of a supervised classifier trained on $\alpha=\#\text{epochs}$ labeled examples.
To quantify performance, we use the \emph{Normalized Mutual Information} (NMI) between its self-generated labels \( \hat{y} \) and the true labels \( y^\star \) used in the main text:
\begin{equation}
\mathrm{NMI} = \frac{I(\hat{y}; y^\star)}{\sqrt{H(\hat{y}) H(y^\star)}},
\end{equation}
where \( I(\cdot;\cdot) \) denotes mutual information and \( H(\cdot) \) the Shannon entropy.
NMI ranges from 0 (no information) to 1 (perfect recovery up to permutation).
For small noise levels, the two approaches achieve essentially the same performance once this rescaling is taken into account. 
At larger $\sigma$, supervised training degrades more rapidly, while the unsupervised agent preserves a more robust representation. 
Interestingly, when the comparison is repeated with a supervised procedure at fixed $\alpha=1$, we find that for intermediate $\sigma$ and sufficiently strong regularization $\lambda$, the turnover dynamics can even slightly outperform the supervised baseline.  

We attribute this effect to the role of self-consistency in the turnover procedure. 
By construction, the unsupervised dynamics avoids direct overfitting of the dataset, since labels are continuously regenerated by the agent itself rather than externally imposed. 
In contrast, supervised learning with strong regularization can still overfit the outlier components of the data (an effect that weakens as $\alpha$ increases), which may explain the relative advantage of the unsupervised approach in certain regimes.

\subsection{Scaling in $\alpha = P/N$}

\begin{figure}
    \centering
    \includegraphics[width=0.5\linewidth]{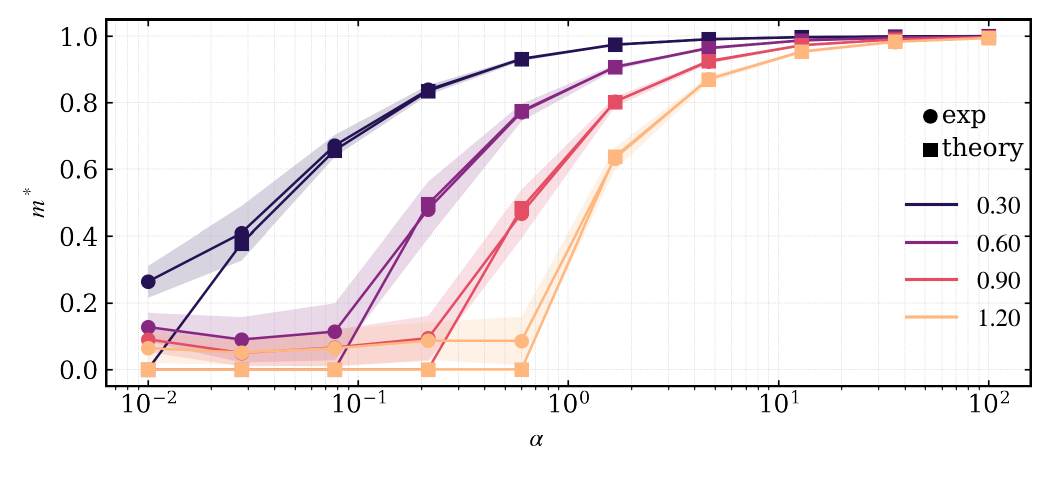}
    \caption{Dependence of the asymptotic overlap $m^\star$ on the effective per-step load $\alpha = P/N$ for normalized weights $m_t = \phi_t$, $q_t = 1$, for different values of the noise parameter $\sigma$ (colors).  
    A sharp transition is observed around $\alpha = O(1)$, with $m^\star$ scaling as a function of $\sigma$.  
    In the small-$\alpha$ regime, the lack of information prevents learning, whereas in the extensive regime $\alpha \sim N$ the system consistently extracts signal regardless of noise level. 
    For small $\alpha$, theoretical predictions may be affected by numerical errors and algorithmic limitations.}
    \label{fig:scaling_alpha}
\end{figure}

We performed numerical experiments to study the evolution of the system’s capability as a function of the effective load $\alpha = P/N$.  
This analysis, reported in Fig.~\ref{fig:scaling_alpha}, reveals a non-trivial yet monotonic scaling.  
In the limit $\alpha \to 0$, only a vanishing fraction of patterns is available and the turnover dynamics fails to extract useful information, leading to negligible values of the order parameter $m^\star = \phi^\star$ (normalized weights $q_t = 1$).  
As $\alpha$ increases, a sharp rise emerges around $\alpha = O(1)$, where the agent becomes sensitive to the signal and the turnover dynamics converge to an informative fixed point with non-zero alignment. 
In this regime, the asymptotic alignment also exhibits a clear dependence on the noise level $\sigma$.  
Finally, when $\alpha$ grows extensively with $N$, the system is supplied with enough information to recover structure at all noise levels.  
In the main text, we focused on the regime $\alpha = O(1)$, as it is both the most delicate and the most relevant for the emergence of learning.

\subsection{Overabundant output dimension}

In the End Matter, we presented results for multi-centroid distributions focused on the case where the output dimension \( K \) matches the number of ground-truth centroids \( C \), i.e., \( K = C \). 
This constraint, however, is not essential. 
In practice, the true number of classes is often unknown or overestimated, making it relevant to study the effect of relaxing this assumption.

To this end, we performed a set of simulations exploring varying levels of output overparameterization, quantified by the surplus dimension \( \Delta = K-C \). 
Our goal was to assess how increasing \( K \) beyond \( C \) affects the emergence of structured representations, particularly the alignment between the agents’ outputs and the underlying class structure.

\begin{figure}
    \centering
    \includegraphics[width=0.75\linewidth]{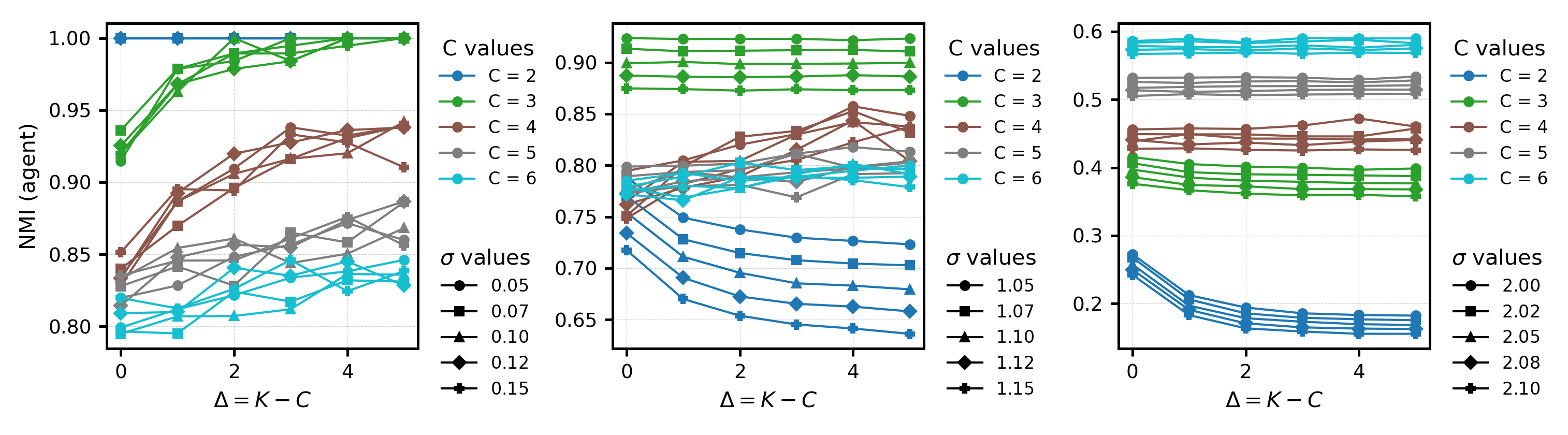}
    \caption{Normalized mutual information (NMI) between agents' outputs and ground-truth class labels as a function of the overparameterization $\Delta = K-C$, where $K$ is the output dimension and $C$ is the number of classes.
    Results are shown across three noise regimes (left to right: low, medium, and high), with the noise amplitude scaled as $\sigma_c = \sigma / C$ and the data-to-dimension ratio fixed as $\alpha_c = \alpha \cdot C$ with $\alpha = 2$.
    Each curve corresponds to a different number of classes $C$ (color), and different noise amplitudes $\sigma$ (marker style).
    In the low-noise regime (left), performance consistently improves with increasing $\Delta$. 
    In the medium-noise regime (center), performance benefits from larger $\Delta$ only for $C > 3$.
    In the high-noise regime (right), performance saturates and becomes largely insensitive to $\Delta$.
 } \label{fig:NMIvsDiff}
\end{figure}

As shown in Fig.~\ref{fig:NMIvsDiff}, the impact of overparameterization depends strongly on the noise regime. 
In the low-noise regime (left panel), increasing \( \Delta \) leads to a clear and consistent improvement in alignment, indicating that additional output dimensions can enhance the agent’s ability to recover the ground-truth structure. 
In contrast, in the high-noise regime (right panel), performance saturates and becomes largely insensitive to \( \Delta \).

\begin{figure}
    \centering
    \includegraphics[width=\linewidth]{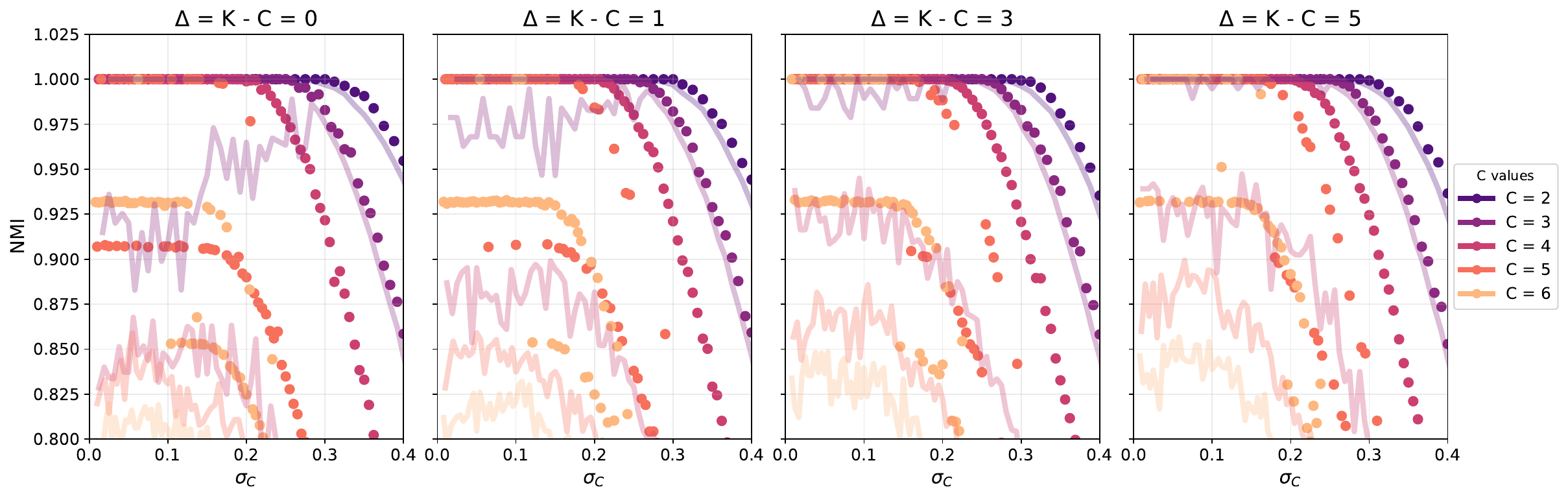}
    \caption{NMI as a function of $\sigma_C$ for different $\Delta = K-C$ and $C$: markers indicate the best NMI achieved across runs, while solid lines show the corresponding averages, highlighting the contrast between possible perfect classifications and the smaller average, brought down by mediocre outcomes.}
    \label{fig:NMIvsSigma_delta}
\end{figure}

This phenomenon is further illustrated in Fig.~\ref{fig:NMIvsSigma_delta}. 
Note that the average NMI, on its own, does not provide a faithful representation of the agent’s behavior: due to the strong dependence on the initial conditions, favorable initializations can occasionally lead to perfect recovery, reaching $\text{NMI} = 1$. 
Most runs, however, fail to achieve such alignment, and the average is instead dominated by more mediocre outcomes. 
In this sense, the average NMI conflates two distinct regimes: rare but highly successful runs and more typical runs that remain far from the optimum.
Increasing the number of latent dimensions $K$ partially mitigates this issue, as additional directions raise the probability that at least a subset aligns with the underlying structure of the data, thereby increasing the likelihood of achieving a good NMI in practice.
Notably, this effect extends over a wider range of $\sigma_C$ values than the region where perfect alignment is possible (where the best possible NMI degrades naturally).

\begin{figure}
    \centering
    \includegraphics[width=0.85\columnwidth]{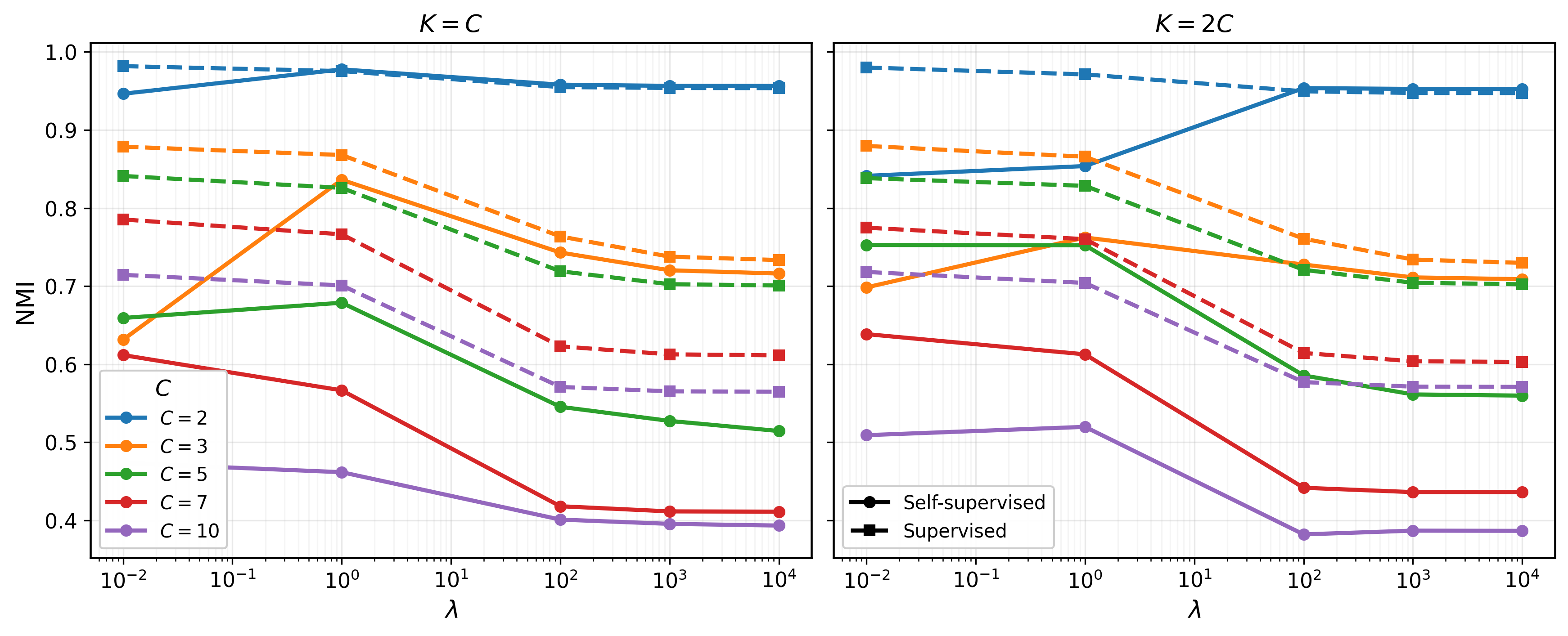}
    \caption{Normalized Mutual Information (NMI) between predicted and ground-truth labels on MNIST as a function of the regularization strength \(\lambda\), for unsupervised turnover dynamics (solid lines, circle markers) and supervised training (dashed lines, square markers).
    Results are shown for \(C \in \{2, 3, 5, 7, 10\}\) digits (colors) in the matched  output setting \(K=C\) (left) and the overparameterized setting \(K=2C\) (right). 
    Each point is averaged over 10 independent runs.
    The supervised baseline is trained on a dataset of the same size as each turnover step. 
    Despite the increased complexity of MNIST with respect to Gaussian mixtures, the unsupervised dynamics reaches non-trivial and often supervised-comparable NMI.
    Both protocols display the same optimal regularization scale, while overparameterization is beneficial only in selected regimes and can be detrimental at low \(C\) and weak regularization.}
    \label{fig:mnist_nmi}
\end{figure}

{
\subsection{Multiclass analysis on MNIST}

To assess whether the multiclass phenomenology observed in synthetic Gaussian mixtures (End Matter) extends to real structured data, we repeated the analysis on the standard MNIST dataset \cite{mnist} using the Normalized Mutual Information (NMI) as a performance metric.
For each value of \(C\), we restricted the dataset to the first \(C\) digits, namely \(0,\dots,C-1\), and compared two output configurations: the matched case \(K=C\) and an overparameterized setting with \(K=2C\). 
The NMI between the predicted labels \(\hat{y}\) and the true digit labels \(y^\star\) provides a permutation-invariant measure of class recovery, ranging from \(0\) for uninformative assignments to \(1\) for perfect reconstruction of the ground-truth partition.

The results, summarized in Fig.~\ref{fig:mnist_nmi}, show that the unsupervised turnover dynamics is able to extract a substantial amount of class information even in this considerably more complex setting. 
Unlike the synthetic Gaussian mixture, MNIST contains strong intra-class variability, non-linear class boundaries, and digit-dependent correlations that are not captured  by a simple centroid-plus-noise description. 
Despite this additional structure, the dynamics reaches high NMI for small and intermediate values of \(C\), and remains quantitatively comparable to the supervised baseline in several regimes. 
As expected, performance decreases as the number of digit classes increases, reflecting the progressive difficulty of stabilizing a coherent multiclass partition. 
Nevertheless, the persistence of non-trivial NMI values even for larger \(C\) indicates that the turnover mechanism is not limited to idealized Gaussian data, but can also recover meaningful class structure in realistic datasets.

A further qualitative difference with respect to the Gaussian-mixture setting is the role of the regularization strength. 
On MNIST, both the unsupervised and supervised curves display an optimal range of \(\lambda\), typically of order one, beyond which performance does not improve and may even degrade. 
This suggests that, for structured real data, regularization does not merely 
stabilize the self-consistent dynamics, but also controls the effective capacity of the classifier in both the supervised and unsupervised protocols.
Too weak regularization leads to unstable or poorly organized representations, whereas too strong regularization limits the ability to exploit the finer informative directions present in the data.

Finally, the effect of output overparameterization is more nuanced than in the Gaussian-mixture experiments. 
Increasing the number of output units to \(K=2C\) can improve the NMI in some regimes, showing that redundant output dimensions may help the dynamics discover informative partitions. 
However, this effect is not uniformly beneficial: in particular, at low \(C\) and small \(\lambda\), overparameterization can reduce performance, presumably because the additional outputs introduce competing or redundant self-generated labels before the dynamics is sufficiently regularized. 
Thus, while overparameterization can still assist class recovery, its benefit on MNIST depends on the balance between task complexity and regularization, rather than acting as a systematic advantage across the whole parameter range.}

\subsection{Learning under data constraints}
We now examine the effect of a finite training budget on the learning dynamics, focusing on how different data usage strategies impact the emergence of structure.

As an initial investigation of replication under such constraints, we consider a fixed total data budget \( P_{\text{tot}} \), and explore how different input sampling strategies affect the emergence of structure.
We analyze three representative protocols:
\begin{enumerate}
    \item \emph{Alternating fixed batches}, where two (or more) disjoint mini-batches are defined once and then reused across successive time steps;  
    \item \emph{Uniform splitting} (or \emph{0\% resampling}), where the budget is partitioned into \( T \) disjoint subsets of size \( P = P_{\text{tot}} / T \), each of which is presented only once;  
    \item \emph{Partial \text{X}\% resampling}, where at each step a fraction \( P / P_{\text{tot}} = \text{X}/100 \) is sampled uniformly at random from the entire dataset, so that examples may be reused across time with a tunable frequency.
\end{enumerate}

\begin{figure}
    \centering
    \includegraphics[width=0.42\linewidth]{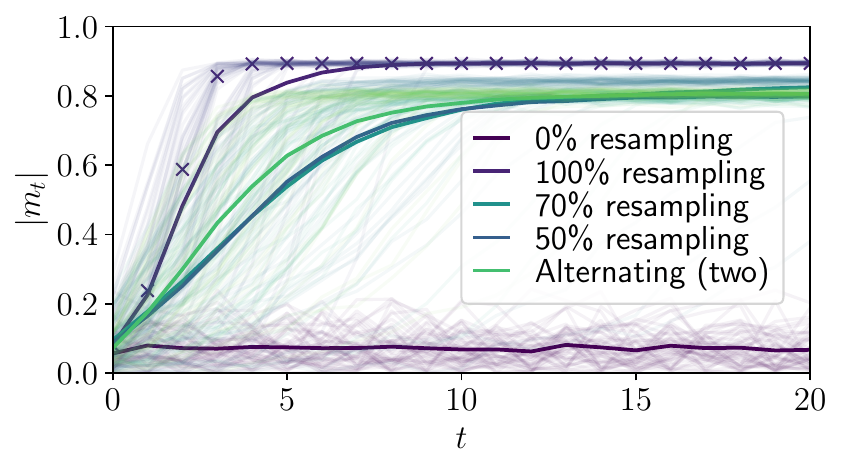}
    \caption{Evolution of the alignment with the signal for a mixture of $C = 2$ Gaussians for different training protocols, averaged over 50 numerical simulations with $\alpha = 1$, $N = 10^3$ and $\sigma = 0.5$, $\lambda =10$; faint lines indicate individual trajectories, markers show the $N\to \infty$ replica solution for the standard protocol starting from the experimental mean alignment at $t=0$.}
    \label{fig:SM_underconstraint}
\end{figure}

The comparison, reported in Fig.~\ref{fig:SM_underconstraint}, reveals that replicators can still emerge under strong data limitations, provided that each step offers sufficiently rich statistics.  
Protocols~1 and~3, although they involve data reuse, preserve enough information to support robust and self-consistent generalization.  
In contrast, the uniform-splitting regime severely limits the information content of each update: since batches are too small, the dynamics typically fail to converge to a structured state.

These results illustrate that, under strict data constraints, the decisive factor is not temporal novelty per se, but the statistical information contained in each update.   
This mechanism may also be relevant in broader settings of decentralized or resource-limited adaptation, where performance depends less on data abundance than on efficient reuse and compression of available information.

\begin{figure}
    \centering
    \includegraphics[width=0.58\linewidth]{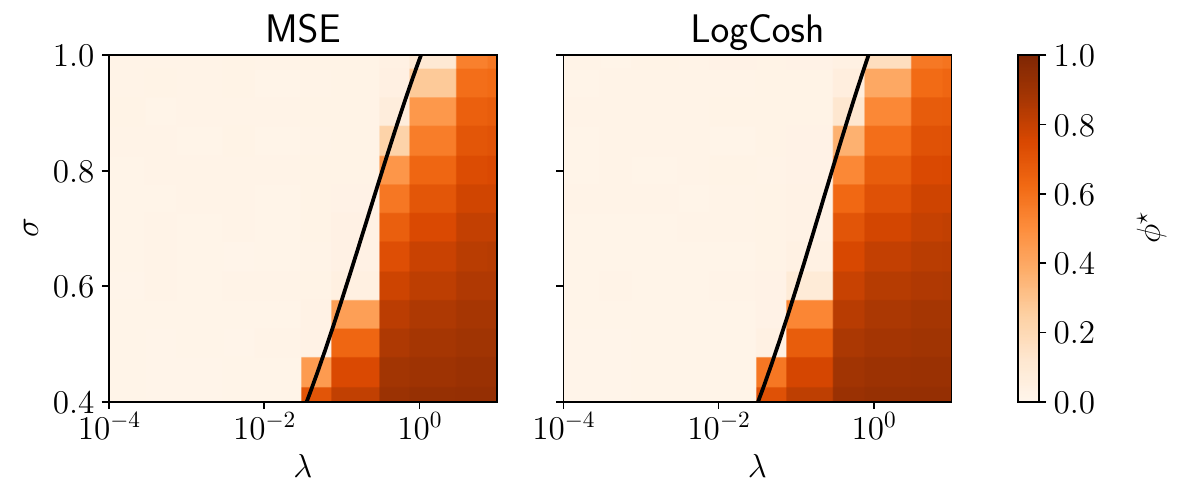}
    \caption{Reproduction of the spontaneous learning phase diagram for other training losses. The theory (continuous line) still displays a very good match with numerical experiments.}
    \label{fig:extra_losses}
\end{figure}

{
\subsection{Other loss functions}

In order to assess the robustness of our results to losses other than the cross-entropy loss that we consider in the main text, we have reproduced results for the markedly different mean-squared error (MSE) and log-cosh (LC) losses,
\begin{equation}
    \ell_\mathrm{MSE}(y,x) = \frac12 (y-x)^2, \qquad \ell_\mathrm{LC}(y,x) = \log \cosh(y-x).
\end{equation}.
Note that while the cross-entropy is a margin loss which compares the label $y$ to a nonlinear function of the margin $x$, these are residual losses, much less adapted to classification tasks. As a result, we observe a significant shift in the phase boundary, see Fig.~\ref{fig:extra_losses} and an overall significant reduction of the region where spontaneous learning occurs. Nonetheless, the phenomenology is perfectly analogous, and high regularization still enables learning. Note also that the fact that these two losses result in very similar phase boundaries is to be expected as their behavior is similar close to zero, $\log \cosh(\varepsilon) = \frac12 \varepsilon^2 + O(\varepsilon^4)$.

Note that the replica theory detailed below is in fact amenable to any arbitrary choice of the loss, the replica-symmetric ansatz simply requiring that the loss is convex in the weights. As shown in the figure, the phase boundary computed using the method described in the main text adapted to the two trial regression losses still provides a very close prediction to numerical experiments.
}

\subsection{Population scaling in the collective setting}

\begin{figure}
    \centering
    \includegraphics[width=.5\linewidth]{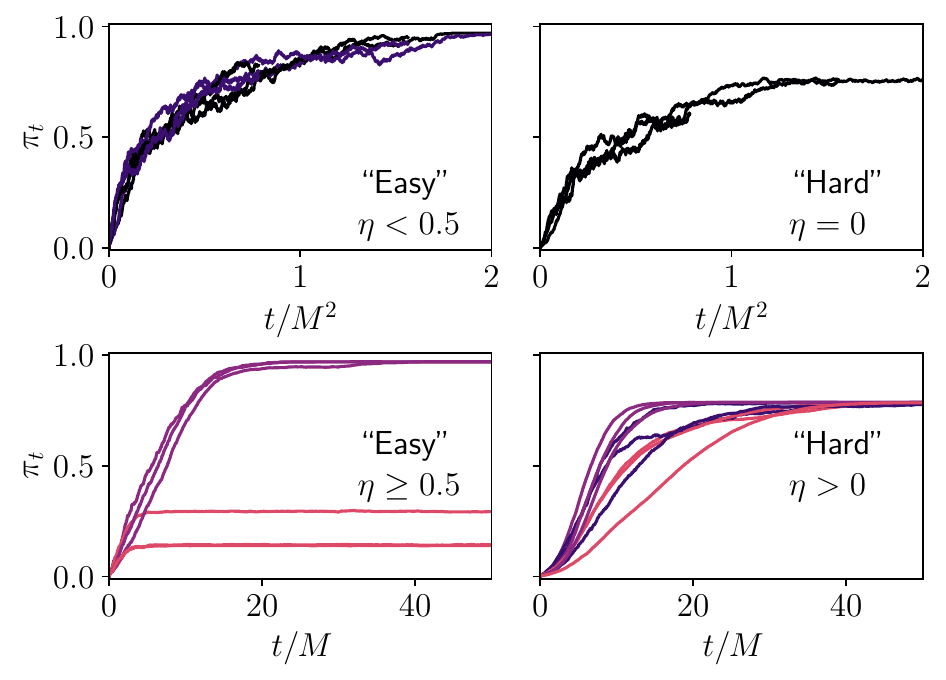}
    \caption{Evolution of the absolute value of the average alignment describing consensus within the population in fully connected populations of size $M = [32,64,128]$, $N = 500$, $\alpha = 1$, $\lambda = 10$, averaged over 32 trials. 
    Colors indicate the value of $\eta$, matching the legend of Fig.~\ref{fig:ABM_experiments}.
    Top: rescaling with the size of the population \textit{squared}, corresponding to fluctuation-driven dynamics. 
    Bottom: standard rescaling with the size of the population, corresponding to aligning dynamics.
    Left: ``Easy'' learning problem, $\sigma = 0.25$.
    Right: ``Difficult'' learning problem, $\sigma = 0.75$. 
    For harder problems, where agents learn simultaneously, only the $\eta = 0$ case remains fluctuation-driven.}
    \label{fig:SM_pop_scaling}
\end{figure}

As discussed in the End Matter, the study of a population of agents reveals a crossover in the evolution of the collective alignment. 
Indeed, when the learning problem is too easy, the agents quickly reach a high individual alignment to $\pm \mu$, followed by a slow, $O(M^2)$, fluctuation-driven formation of consensus. 
When the problem is harder, learning and consensus formation evolve in tandem, leading to full consensus in a much faster $O(M)$ scaling. 
To illustrate these two distinct temporal dynamics, we show numerical simulations for different values of $M$ in the regions of interest in Fig.~\ref{fig:SM_pop_scaling}.

\section{Detailed calculations}

\subsection{Long-time correlations and stability} 

\begin{figure}
    \centering
    \includegraphics[width=0.8\linewidth]{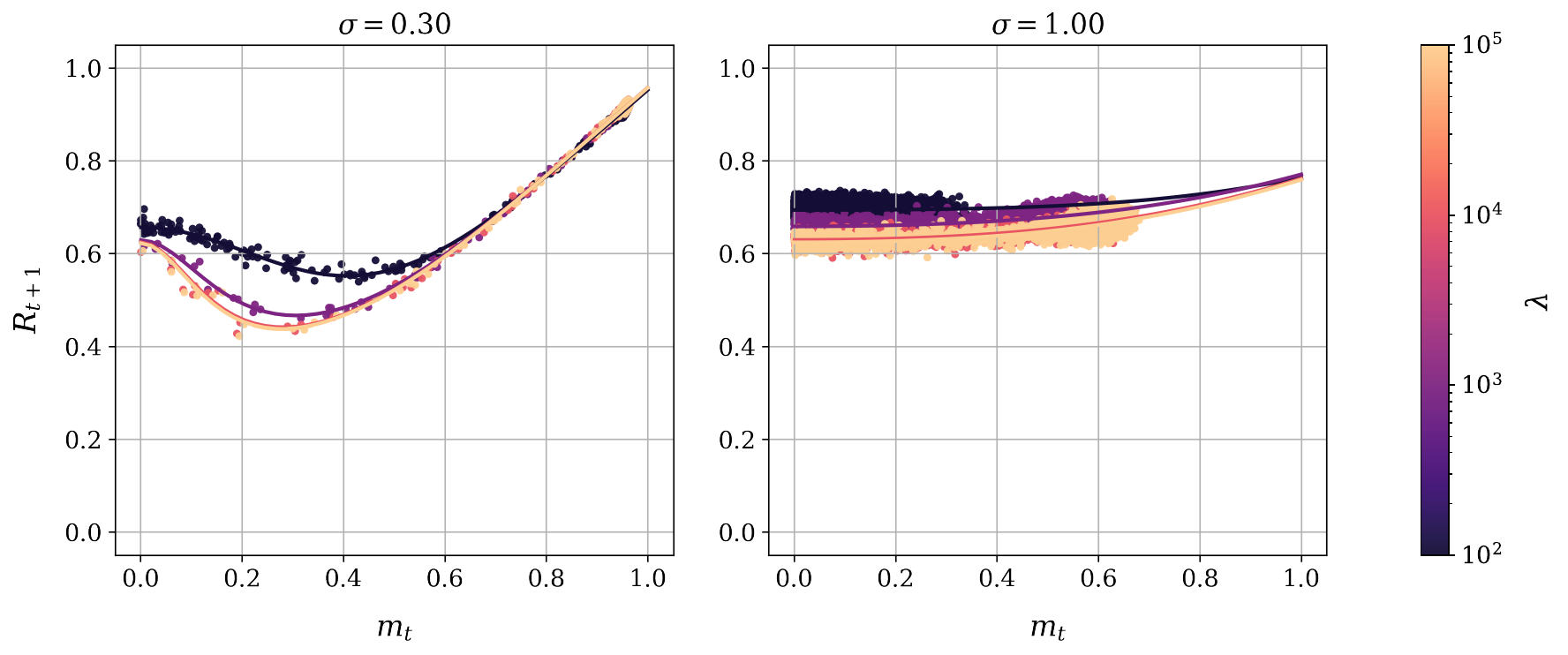}
    \caption{%
    Alignment $R_{t+1}$ versus centroid overlap $m_t$ at fixed $\sigma$ (left: $0.30$; right: $1.00$). 
    Dots are experimental data colored by $\lambda$ (log scale); solid lines are theoretical predictions. 
    Results are averaged over 50 independent runs with $N=10^3$, $\alpha=1$. 
    The overlap remains systematically large ($R_{t+1}\gtrsim0.4$) even when the dynamics do not converge to any fixed point, illustrating that $R$ is a poor indicator of long-term stability.}
    \label{fig:QvsM}
\end{figure}

A natural candidate to quantify dynamical stability is the overlap between consecutive configurations, defined as
\begin{equation}
    R_t = \frac{\bm{w}_t \cdot \bm{w}_{t+1}}{\|\bm{w}_t\| \, \|\bm{w}_{t+1}\|}.
\end{equation}
However, this measure can be misleading for two reasons.  
(i) By construction, consecutive configurations are correlated due to the update rule. 
As a result, $R_t$ remains far from zero even in regimes where the dynamics do not converge to a stable functional state (as shown in Fig.~\ref{fig:QvsM}), making it a poor indicator of long-term stability. 
(ii) Since the dynamics is intrinsically stochastic, even a high value of $R_t<1$ does not guarantee stability: small random perturbations at each step may accumulate and drive the system far from its initial configuration.  

A more informative observable is the overlap of the weights with a fixed external reference vector, such as the centroid $\mu$ of the data distribution. 
If this overlap converges to a stationary value, the system can be regarded as dynamically stable. 
Assuming that the only stable directions are those associated with the centroids, we can establish a lower bound on the long-time correlation
\begin{equation}
    C(\Delta t) = \frac{\bm{w}_t \cdot \bm{w}_{t+\Delta t}}
    {\|\bm{w}_t\| \, \|\bm{w}_{t+\Delta t}\|}.
\end{equation}
In particular, it follows that:

\medskip
\begin{enumerate}
    \item \emph{Lower bound.}  
    Decompose the normalized weight vector as $\frac{\bm{w}_t}{\|\bm{w}_t\|} = \phi_t \bm{\mu} + \bm{\xi}_t$, with $\langle \bm{\mu} \cdot \bm{\xi}_t\rangle = 0$.  
    For a slow dynamical process it is natural to assume that fluctuations retain some positive correlation across nearby times (i.e.\ $\langle \bm{\xi}_t \cdot \bm{\xi}_{t+1} \rangle > 0$), which is consistent with our learning dynamics.  
    Then
    \[
    C(\Delta t) =  \phi_t \phi_{t+\Delta t} + \langle \bm{\xi}_t \cdot \bm{\xi}_{t+\Delta t}\rangle.
    \]
    The first term tends to $(\phi^\star)^2$, while the second is non-negative under the above assumption, obtaining $C(\Delta t)\ge (\phi^\star)^2$, where $\phi^\star$ denotes the nontrivial fixed point obtained in the stability analysis of the main text.

    \item \emph{Long-time limit.}
    Under dynamical stability, $\phi_t \to \phi^\star$.
    Meanwhile, the orthogonal components $\bm{\xi}_t$ evolve stochastically within the high-dimensional subspace orthogonal to the ground truth direction $\bm{\mu}$.
    In the long-time limit, these fluctuations decorrelate, so that 
    $
    \lim_{\Delta t \to \infty} 
    \langle \bm{\xi}_t \cdot \bm{\xi}_{t+\Delta t} \rangle = 0
    $.
    Hence the correlation reduces to
    \[
    \lim_{\Delta t\to\infty} C(\Delta t) = (\phi^\star)^2.
    \]
    Intuitively, the only persistent contribution to alignment is the projection along the stable centroid direction.
\end{enumerate}

\begin{figure}
    \centering
    \includegraphics[width=\linewidth]{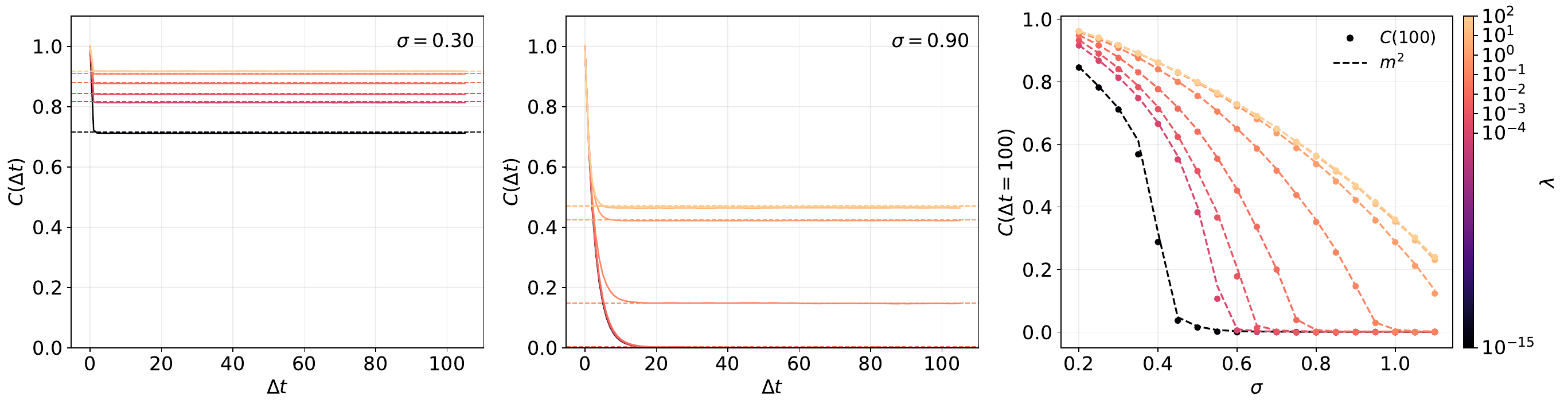}
    \caption{%
    Temporal correlation of the agent’s weights. 
    (a) Evolution of the overlap $C(\Delta t)$ for $\sigma=0.30$ and different values of the regularization strength $\lambda$. 
    (b) Same as in (a) but for $\sigma=0.90$. 
    (c) Comparison between the long-time correlation $C(100)$ (symbols) and the theoretical prediction $(m^\star)^2$ (dashed lines), as a function of the noise level $\sigma$ for varying $\lambda$. 
    Results are averaged over 50 independent runs with $N=10^3$, $\alpha=1$. 
    The agreement confirms that $C(\Delta t) \to (m^\star)^2$ at long times, validating the identification of functional stability with persistent alignment to the centroid.}
    \label{fig:C(100)}
\end{figure}

In Fig.~\ref{fig:C(100)} we verify this prediction numerically, finding excellent agreement. 
This justifies our interpretation of functional stability as the emergence of persistent alignment between the agent and the centroid direction.

\clearpage

\subsection{Replica computation} \label{sec:replica}

In this section, we use the replica method to track the evolution of the overlap
\[
  \phi_t = \,\frac{{\bm{w}}_t \cdot {{\bm{\mu}}}}
                   {\lVert \bm{w}_t\rVert},
\]
i.e.\ the correlation between the agent’s weight vector $\bm{w}_t$ and the centroid vector ${\bm{\mu}}$, adapting the procedure outlined in \cite{SaraoMannelli2024Bias} to our setting.

We detail the computation in the case where the data is sampled from $C=2$ centroids. While it can be generalized to $C > 2$, the computation is lengthy and the resulting saddle-point equations difficult to solve numerically \cite{Loureiro2021GMM}, while not being essential for our study. The case of interest is the one in which the agent, at the iteration $t+1$, can only observe its labels generated at the previous time step, with the weights $\W_t$.
By defining a loss function $\ell$, we want to construct the weights $\W_{t+1}$ that minimize the loss, with the previous labels.
For $P=\alpha N$, the loss of interest is defined as:
\begin{equation}
\mathcal{L}_t \left(\W_{t},\W_{t+1}, \{x^\nu\}; \lambda\right) = \sum_{\nu=1}^{\alpha N} \ell \left(\sign \left(\frac{\W_{t}\cdot x^{\nu}}{\sqrt{N}}\right),\,\frac{\W_{t+1} \cdot x^{\nu}}{\sqrt{N}}\right) + \frac{\lambda}{2} \left\lVert \W_{t+1} \right\rVert^2,
\label{eq:loss}
\end{equation}
where $\sign(\cdot)$ is the sign function, and we have removed the time index where possible to lighten the notation.
Here, the function $\ell(y, x) = -\log h(y x) = \log(1 + \e^{-yx})$ is the logistic loss, with $h(z) = 1/(1+\e^{-z})$ and $y = \pm1$ a hard label. As such, it is identical to the expression of the main text, Eq.~\eqref{eq:crossentropy}.

The key quantity to be computed to address the values of the overlap parameters of interest is the free entropy of the model:
\begin{equation}
\begin{aligned}
\Phi &= \lim_{\beta \to \infty} \frac{1}{\beta N} \left\langle \log \int \dd \W_{t+1} \,\e^{-\beta \mathcal{L}_t(\W_{t+1},\W_t,\{x^\nu\};\lambda)} \right\rangle_{\{\bm{x}^\nu\},\W_t} \\
&= \lim_{\beta \to \infty} \frac{1}{\beta N}\left\langle Z_{t}^{-1}\int \dd\mu \left(\W_t\right)\log\int \dd\mu(\W_{t+1})\int \prod_{\nu=1}^{\alpha N}\e^{-\beta\,\ell\left(\sign\left(\frac{\W_{t}\cdot x^{\nu}}{\sqrt{N}}\right),\,\frac{\W_{t+1} \cdot x^{\nu}}{\sqrt{N}}\right)}\right\rangle_{\{\bm{x}^\nu\}},
\end{aligned}
\label{eq:freeentropy2centroids}
\end{equation}
where $\mu(\W_t)$ is the un-normalized distribution of the previous iteration weights and $Z_t$ is its normalization.

\subsubsection{Teacher partition function}

The first step of the computation consists in obtaining the analytical expression for the teacher partition function $Z_t$. As mentioned in the main text, the labeling does not depend on the norm of the weights. We can therefore, without loss of generality, fix the normalization of the teacher to $q_t = 1$, and set its overlap to the ground truth to $m_t = \frac{1}{N} \W_t \cdot \mu = \phi_t$,
\begin{equation}
Z_{t}=\int \dd\mu \left(\W_t\right) = \int \dd \W_t \, \delta \left(\left\|\W_t\right\|^{2}-N\right)\delta \left(\W_t\cdot \mu-N \phi_t\right).
\end{equation}
Using the Fourier representation of the Dirac $\delta$,
\begin{equation*}
\delta\left(x-aN\right) = \int\frac{\dd\hat{a}}{2\pi/N}\, \e^{-i\hat{a}\left(\frac{x}{N}-a\right)},
\end{equation*}
one obtains
\begin{equation}
Z_t = \int \frac{\dd\hat{q}_t}{2\pi/N}\, \frac{\dd\hat{\phi}_t}{2\pi/N}\, \e^{N (g_I^{(t)} + g_S^{(t)})},
\end{equation}
with
\begin{align}
g_I^{(t)} &= - \hat{q}_t - \hat{\phi}_t \phi_t,\\
g_S^{(t)} &= \frac{\hat{\phi}^2_t}{2(1-2\hat{q}_t)} - \frac{1}{2} \log(1-2\hat{q}_t).
\end{align}
In the $N\to \infty$ limit, the integral is dominated by the saddle point of the exponent,
\begin{equation}
Z_t \sim \exp \left( - \hat{q}_t - \hat{\phi}_t \phi_t + \frac{\hat{\phi}^2_t}{2(1-2\hat{q}_t)} - \frac{1}{2} \log(1-2\hat{q}_t) \right)^N = z_t^N,
\label{eq:Zt}
\end{equation}
where the conjugate variables satisfy the saddle-point equations
\begin{equation}
\hat{\phi}_t = -\frac{\phi_t}{\phi_t^2 - 1}, \qquad \hat{q}_t = \frac{\phi_t^2}{2(\phi_t^2 - 1)}, \qquad \text{equivalently} \qquad 1-2\hat{q}_t = \frac{1}{1-\phi_t^2}, \qquad \phi_t = \frac{\hat{\phi}_t}{1-2\hat{q}_t}.
\label{eq:teacher_saddle}
\end{equation}

\subsubsection{Replicated student partition function}

Having expressed the measure of the teacher weights as a function of the alignment $\phi_t$, we employ the replica trick
\begin{equation}
\log Z_{t+1} = \lim_{n\to0} \frac{Z_{t+1}^n-1}{n},
\end{equation}
to compute the average of the logarithm of the updated weights' partition function, which reads
\begin{equation}
Z_{t+1}^{n}=Z_{t}^{-1}\int \dd\mu \left(\W_t\right)\int\prod_{a}\dd\mu \left(\W_{t+1}^{a}\right)\left\langle \prod_{\nu=1}^{\alpha N}\prod_{a}\e^{-\beta\,\ell\left(\sign\left(\frac{\W_{t}\cdot x^{\nu}}{\sqrt{N}}\right),\,\frac{\W_{t+1}^a\cdot x^{\nu}}{\sqrt{N}}\right)}\right\rangle _{\{\bm{x}^\nu\}}.
\end{equation}

It is now possible to perform the expectation over the realization of the dataset (quenched disorder).
The first step requires the changes of variables $u_\nu = \frac{\W_{t}\cdot x^{\nu}}{\sqrt{N}}$ and $\xi_\nu^a = \frac{\W_{t+1}^a\cdot x^{\nu}}{\sqrt{N}}$.
By performing these changes of variable, the replicated partition function becomes:
\begin{align}
Z_{t+1}^n &= \Bigg\langle Z_{t}^{-1}\int \dd\mu \left(\W_{t}\right)\int\prod_{a}\dd\mu \left(\W_{t+1}^{a}\right)\int\prod_{\nu=1}^{\alpha N}\frac{\dd u_{\nu}\dd\hat{u}_{\nu}}{2\pi}\,\e^{i\hat{u}_{\nu}\left(u_{\nu}-\frac{\W_{t}\cdot x^{\nu}}{\sqrt{N}}\right)}\nonumber\\
&\hspace{1.5em}\times\int\prod_{\nu=1}^{\alpha N}\prod_{a}\frac{\dd\xi_{\nu}^{a}\dd\hat{\xi}_{\nu}^{a}}{2\pi}\,\e^{i\hat{\xi}_{\nu}^{a}\left(\xi_{\nu}^{a}-\frac{\W_{t+1}^a\cdot x^{\nu}}{\sqrt{N}}\right)}\prod_{a}\prod_{\nu=1}^{\alpha N}\e^{-\beta\,\ell\left(\sign(u_{\nu}),\,\xi_{\nu}^{a}\right)}\Bigg\rangle_{\{\bm{x}^\nu\}}.
\end{align}

Performing the disorder average, with $x^\nu = c^\nu \mu/\sqrt{N} + z^\nu$ and $z^\nu \sim \mathcal{N}(0,\sigma^2/N)$, one obtains
\begin{align}
\Bigg\langle &\e^{-i\sum_{a}\hat{\xi}_{a}^{\nu}\frac{x^{\nu}\cdot\W_{t+1}^{a}}{\sqrt{N}}-i\hat{u}_{\nu}\frac{\W_{t}\cdot x^{\nu}}{\sqrt{N}}} \Bigg\rangle \nonumber\\
&=\,\e^{-ic^{\nu}\left(\sum_{a}\hat{\xi}_{a}^{\nu}\frac{\sum_{i}\W_{t+1,i}^{a}\mu_{i}}{N}+\hat{u}^{\nu}\frac{\sum_i \W_{t,i}\mu_{i}}{N}\right)}\e^{-\frac{\sigma^{2}}{2}\left(\sum_{ab}\hat{\xi}_{a}^{\nu}\hat{\xi}_{b}^{\nu}\frac{\sum_{i}\W_{t+1,i}^{a}\W_{t+1,i}^{b}}{N}+2\hat{u}^{\nu}\sum_{a}\hat{\xi}_{a}^{\nu}\frac{\sum_{i}\W_{t+1,i}^{a}\W_{t,i}}{N}+\left(\hat{u}^{\nu}\right)^{2}\frac{\sum_i \W_{t,i}^2}{N}\right)},
\end{align}
where $c^\nu = \pm 1$ is the cluster assignment.
Once the data generative process is known, we introduce the natural overlaps:
\begin{align}
m^{a}_{t+1}&=\frac{\sum_i \W_{t+1,i}^{a}\mu_{i}}{N}, \qquad
q^{ab}_{t+1}=\frac{\sum_{i}\W_{t+1,i}^{a}\W_{t+1,i}^{b}}{N}, \qquad
R_t^{a}=\frac{\sum_{i}\W_{t+1,i}^{a}\W_{t,i}}{N},\\
m_t&=\frac{\sum_i \W_{t,i}\mu_{i}}{N}, \qquad q_t=\frac{\sum_{i}\W_{t,i}^2}{N}.
\end{align}
Without loss of generality, we fix $q_t = 1$, so that $m_t = \phi_t$ just as in the student partition function computation.
Introducing these overlaps via Lagrange multipliers, the replicated partition function factorizes and takes the form
\begin{equation}
Z_{t+1}^{n}=\int \frac{\dd\hat{q}_t}{2\pi/N}\, \frac{\dd\hat{\phi}_t}{2\pi/N} \int\prod_{a}\frac{\dd m_{t+1}^{a}\dd\hat{m}_{t+1}^{a}}{2\pi/N}\int\prod_{a}\frac{\dd R_t^{a}\dd\hat{R}_t^{a}}{2\pi/N}\int\prod_{ab}\frac{\dd q_{t+1}^{ab}\dd\hat{q}_{t+1}^{ab}}{2\pi/N}\left(G_I\, G_{S}\right)^{N}\left(G_{E}\right)^{\alpha N},
\end{equation}
with
\begin{align}
G_{I}&=\exp \left(-\sum_{a}\hat{m}_{t+1}^{a}m_{t+1}^{a}-\sum_{ab}\hat{q}_{t+1}^{ab}q_{t+1}^{ab}-\sum_{a}\hat{R}_t^{a}R_t^{a}\right),\\
G_{S}&=z_t^{-1}\int\frac{\dd w_t}{\sqrt{2\pi}}\,\e^{-\frac{1}{2}(1-2\hat{q}_t) w_t^{2}+\hat{\phi}_t w_t}\int\prod_{a}\dd\mu \left(w_{t+1}^{a}\right)\nonumber\\
&\quad\times\exp \left(\sum_{a}\hat{m}_{t+1}^{a} w_{t+1}^{a}+\sum_{ab}\hat{q}^{ab}_{t+1} w_{t+1}^{a} w_{t+1}^{b}+\sum_{a}\hat{R}_{t}^{a} w_{t+1}^{a} w_t\right),\\
G_{E}&=\frac{1}{2}\sum_{c=\pm 1}\int\frac{\dd u \,\dd\hat{u}}{2\pi}\,\e^{iu\hat{u}}\int\prod_{a}\left(\frac{\dd\xi^{a}\dd\hat{\xi}^{a}}{2\pi}\,\e^{i\xi^{a}\hat{\xi}^{a}}\right)\nonumber\\
&\quad\times\e^{-\frac{\sigma^{2}}{2}\sum_{ab}\hat{\xi}_{a}\hat{\xi}_{b}q_{t+1}^{ab}-\sigma^{2}\hat{u}\sum_{a}\hat{\xi}_{a}R_t^{a}-\frac{\sigma^{2}}{2}\hat{u}^{2}}\prod_{a}\e^{-\beta\,\ell\left(\sign(u+c\phi_t),\,\xi^{a}+c\,m_{t+1}^{a}\right)}.
\end{align}
Importantly, note that the teacher-related multipliers $\hat{q}_t$ and $\hat{\phi}_t$ differ in nature from the student conjugates, as they enforce the \textit{prescribed} teacher moments that are fixed inputs. As such, their saddle-point values are those of the teacher partition function evaluated above, independent of the student, i.e. the teacher is a quenched disorder with a fixed distribution, and the student cannot alter the multipliers that pin its moments.

\subsubsection{Replica-symmetric ansatz and zero-temperature scaling}

As is standard in replica theory, we assume a parametric ansatz for the order parameters. We adopt the replica-symmetric (RS) ansatz \cite{Engel2001StatMech}, justified by the convexity of the cross-entropy loss:
\begin{align}
m_{t+1}^{a}&=m_{t+1}, \qquad R_t^{a}=R_t, \qquad \forall a=1,\dots,n, \\
q_{t+1}^{ab}&=q_{t+1} \quad \forall a\neq b, \qquad q_{t+1}^{aa}=Q_{t+1}.
\end{align}
Under this ansatz and taking the $n\to 0$ limit, the interaction term reads
\begin{equation}
G_{I}=\exp \left(-n\left(\hat{m}_{t+1}m_{t+1}+\hat{R}_tR_t+\frac{1}{2}\hat{Q}_{t+1}Q_{t+1}+\frac{n-1}{2}\hat{q}_{t+1}q_{t+1}\right)\right).
\end{equation}
We are interested in the $\beta\to\infty$ limit. To obtain a finite free entropy we follow \cite{SaraoMannelli2024Bias} and rescale:
\begin{equation}
\left(Q_{t+1}-q_{t+1}\right)=\delta q_{t+1}/\beta,\quad \left(\hat{Q}_{t+1}-\hat{q}_{t+1}\right)=-\beta\delta\hat{q}_{t+1},\quad \hat{q}_{t+1}\sim\beta^{2}\hat{q}_{t+1},\quad \hat{m}_{t+1}\sim\beta\hat{m}_{t+1},\quad \hat{R}_t\sim\beta\hat{R}_t.
\end{equation}
Inserting these into the interaction term gives
\begin{equation}
g_I = \lim_{n\to 0}\frac{\log G_I}{n} = -\beta\left(\hat{m}_{t+1}m_{t+1}+\hat{R}_tR_t+\frac{1}{2} \left(\hat{q}_{t+1}\delta q_{t+1}-\delta\hat{q}_{t+1}q_{t+1}\right)\right).
\end{equation}

We now consider the entropic term. Carrying out the Hubbard--Stratonovich decoupling of the quadratic form $\hat{q}_{t+1}\bigl(\sum_a w_{t+1}^a\bigr)^2$, we can write
\begin{align}
G_{S} = z_t^{-1}&\int\frac{\dd w_t}{\sqrt{2\pi}}\,\e^{-\frac{1}{2}(1-2\hat{q}_t) w_t^{2}+\hat{\phi}_t w_t}\\
&\times\int\mathcal{D}z\left[\int\dd\mu \left(w_{t+1}\right)\,\e^{\frac{1}{2}(\hat{Q}_{t+1}-\hat{q}_{t+1}) w_{t+1}^{2}+\left(\sqrt{\hat{q}_{t+1}} z+\hat{R}_{t} w_t+\hat{m}_{t+1}\right) w_{t+1}}\right]^{n},
\end{align}
where $\mathcal{D}z\, = \dd z\, \e^{-z^2/2}/\sqrt{2\pi}$ denotes the standard Gaussian measure and $\dd\mu(w_{t+1}) = \dd w_{t+1}\,\e^{-\frac{\lambda}{2} w_{t+1}^2}/\sqrt{2\pi}$ is the regularization-induced Gaussian measure on the student weights. 
In what follows we will see that the $z_t^{-1}$ prefactor naturally cancels the teacher-only contributions generated by the $w_t$ integration.

The first step is to complete the square selecting only the purely $w_t$ dependent terms, which leads to the change of variable
\begin{equation}
    w_t = \hat{\phi}_t/(1-2\hat{q}_t) + v_t/\sqrt{1-2\hat{q}_t},
\end{equation}
such that 
\begin{equation}
    \frac{\dd w_t}{\sqrt{2\pi}} \, \e^{-\frac{1}{2}(1-2\hat{q}_t) w_t^{2}+\hat{\phi}_t w_t} = \frac{\e^{-\frac12 \frac{\hat{\phi}_t^2}{1-2\hat{q}_t}}}{\sqrt{1-2\hat{q}_t}}\frac{\dd v_t}{\sqrt{2\pi}} \, \e^{-\frac12 v_t^2} = z_t \,\mathcal{D}v_t,
\end{equation}
where the last step comes from the fact that the teacher auxiliary variables are again \textit{fixed} by the teacher partition function saddle point. Therefore it also follows that 
\begin{equation}
    \hat{R}_t w_t = \frac{\hat{R}_t \hat{\phi}_t}{1-2\hat{q}_t} + \frac{\hat{R}_t v_t}{\sqrt{1-2\hat{q}_t}} = \phi_t \hat{R}_t + \frac{\hat{R}_t v_t}{\sqrt{1-2\hat{q}_t}},
\end{equation}
which can naturally be interpreted as a mean plus Gaussian fluctuation structure for the coupling between the student and the teacher. Taking the limit $n \to 0$ and applying the inverse-temperature rescalings, we thus obtain
\begin{equation}
g_{S}=\lim_{n\to 0}\frac{\log G_S}{n}=\int\mathcal{D}z\int\mathcal{D}v_t\,\log\int\frac{\dd w_{t+1}}{\sqrt{2\pi}}\,\e^{\beta\left(-\frac{\lambda+\delta\hat{q}_{t+1}}{2}w_{t+1}^{2}+\left(\sqrt{\hat{q}_{t+1}} z+\frac{\hat{R}_{t}v_t}{\sqrt{1-2\hat{q}_{t}}}+\hat{R}_{t}\phi_{t}+\hat{m}_{t+1}\right)w_{t+1}\right)}.
\end{equation}
In the $\beta\to\infty$ limit the inner Gaussian integral is dominated by its saddle point, yielding
\begin{equation}
g_{S}=\int\mathcal{D}z\int\mathcal{D}v_t\,\frac{\left(\sqrt{\hat{q}_{t+1}} z+\dfrac{\hat{R}_{t}v_t}{\sqrt{1-2\hat{q}_{t}}}+\hat{R}_{t}\phi_{t}+\hat{m}_{t+1}\right)^{2}}{2(\lambda+\delta\hat{q}_{t+1})}.
\end{equation}
Performing the two Gaussian integrals ($\int\mathcal{D}z\, z = 0$, $\int\mathcal{D}z\, z^{2} = 1$, likewise for $v_t$), the cross terms vanish and we are left with
\begin{equation}
g_{S}=\frac{\left(\hat{R}_{t}\phi_{t}+\hat{m}_{t+1}\right)^{2}+\hat{q}_{t+1}+\dfrac{\hat{R}_{t}^{2}}{1-2\hat{q}_{t}}}{2(\lambda+\delta\hat{q}_{t+1})}.
\end{equation}
Finally, the residual $1/(1-2\hat{q}_t)$ factor multiplying $\hat{R}_t^2$ can be rearranged using the teacher saddle-point relation $1-2\hat{q}_t = 1/(1-\phi_t^2)$ from Eq.~\eqref{eq:teacher_saddle}. Substituting this in gives the compact result
\begin{equation}
g_{S}=\frac{\left(\hat{R}_{t}\phi_{t}+\hat{m}_{t+1}\right)^{2}+\hat{q}_{t+1}+\hat{R}_{t}^{2}\left(1-\phi_{t}^{2}\right)}{2(\lambda+\delta\hat{q}_{t+1})}.
\end{equation}

The energetic term reads
\begin{align}
G_{E} &= \frac{1}{2}\sum_{c=\pm 1}\int\prod_{a} \left(\frac{\dd\xi^{a}\dd\hat{\xi}^{a}}{2\pi}\,\e^{i\xi^{a}\hat{\xi}^{a}}\right)\e^{-\frac{\sigma^{2}}{2}\sum_{ab}\hat{\xi}_{a}\hat{\xi}_{b}q_{t+1}^{ab}}\nonumber\\
&\quad\times\int\frac{\dd u\,\dd\hat{u}}{2\pi}\,\e^{iu\hat{u}}\,\e^{-\sigma^{2}\hat{u}\sum_{a}\hat{\xi}_{a}R_t^{a}-\frac{\sigma^{2}}{2}\hat{u}^{2}}\prod_{a}\e^{-\beta\,\ell\left(\sign(u+c\phi_t),\,\xi^{a}+c\,m_{t+1}^{a}\right)}.
\end{align}
Completing the square in $\hat{u}$ and integrating yields a Gaussian factor in $u$,
\begin{equation}
G_{E} = \frac{1}{2}\sum_{c=\pm 1}\int\prod_{a} \left(\frac{\dd\xi^{a}\dd\hat{\xi}^{a}}{2\pi}\,\e^{i\xi^{a}\hat{\xi}^{a}}\right)\e^{-\frac{\sigma^{2}}{2}\sum_{ab}\hat{\xi}_{a}\hat{\xi}_{b}q_{t+1}^{ab}}\int\frac{\dd u}{\sqrt{2\pi}}\,\e^{-\frac{1}{2}\left(u/\sigma - i\sigma\sum_{a}\hat{\xi}_{a}R_t^{a}\right)^{2}}\prod_{a}\e^{-\beta\,\ell\left(\sign(u+c\phi_t),\,\xi^{a}+c\,m_{t+1}^{a}\right)}.
\end{equation}
Imposing the RS ansatz, rescaling the conjugate variables and taking $n\to 0$ followed by $\beta\to\infty$, the $u$-integral produces a Gaussian factor over the teacher pre-activation whose sign determines the label, while the $\xi^a$-integrals collapse onto the inner saddle point over a single auxiliary $w$. The result reads
\begin{equation}
g_{E} = \frac{1}{2}\sum_{c=\pm 1}\sum_{y=\pm 1}\int\mathcal{D}zH \left(-y\,\frac{c\,\phi_t\sqrt{q_{t+1}}+\sigma R_t z}{\sigma\sqrt{q_{t+1}-R_t^{2}}}\right)\mathcal{M}(c,z;y),
\end{equation}

with
\begin{equation}
\mathcal{M}(c,z;y) = \max_{w} \left\{-\frac{w^{2}}{2}-\ell \left(y,\sigma\sqrt{\delta q_{t+1}} w + cm_{t+1}+\sigma\sqrt{q_{t+1}} z\right)\right\},
\end{equation}
and $H(x) = \int_{x}^{\infty}\mathcal{D}z$. The reduction of the replicated free entropy to a static, low-dimensional problem rests on the assumption of fresh data at each iteration; correlations across steps would require the framework of \cite{takanami2025effect,takahashi2022role}.

\subsubsection{Saddle-point equations}

Collecting the three contributions, the free entropy density at step $t+1$ reads
\begin{equation}
\Phi_{t+1} = \operatorname*{extr}\,\bigl[g_I + g_S + \alpha\,g_E\bigr],
\end{equation}
where the extremization is over both physical and conjugate order parameters. Suppressing the $t+1$ subscript on the student variables ($m\equiv m_{t+1}$, $q\equiv q_{t+1}$, $\delta q \equiv \delta q_{t+1}$, $R \equiv R_t$) and writing $\phi \equiv \phi_t$,
\begin{align}
g_I &= - \left[\hat{m}\,m + \hat{R}\,R + \frac{1}{2} \left(\hat{q}\,\delta q - \delta\hat{q}\,q\right)\right],\\
g_S &= \frac{(\hat{m}+\phi\hat{R})^{2} + \hat{q} + (1-\phi^{2})\hat{R}^{2}}{2(\lambda+\delta\hat{q})},\\
g_E &= \frac{1}{2}\sum_{c=\pm 1}\int\mathcal{D}z\,\Bigl[H(-A_c - Bz)\,\mathcal{M}(c,z;+1)+ H(A_c + Bz)\,\mathcal{M}(c,z;-1)\Bigr],
\end{align}
with auxiliary fields
\begin{equation}
A_c = \frac{c\,\phi\sqrt{q}}{\sigma\sqrt{q-R^{2}}}, \qquad B = \frac{R}{\sqrt{q-R^{2}}},
\end{equation}
and
\begin{equation}
\mathcal{M}(c,z;y) = \max_{w} \left\{-\frac{w^{2}}{2}-\ell \left(y,\sigma\sqrt{\delta q}\,w + c\,m + \sigma\sqrt{q} z\right)\right\}.
\end{equation}

Extremizing first over the conjugate order parameters in $g_I$ eliminates $\hat{m},\hat{R},\hat{q},\delta\hat{q}$ as Lagrange multipliers. Differentiating $g_S$ at fixed conjugates gives the \emph{primal} relations
\begin{equation}
m = \frac{\hat{m}+\phi\hat{R}}{\lambda+\delta\hat{q}},\qquad R = \frac{\phi(\hat{m}+\phi\hat{R})+(1-\phi^{2})\hat{R}}{\lambda+\delta\hat{q}},\qquad q = \frac{(\hat{m}+\phi\hat{R})^{2}+\hat{q}+(1-\phi^{2})\hat{R}^{2}}{(\lambda+\delta\hat{q})^{2}},\qquad \delta q = \frac{1}{\lambda+\delta\hat{q}},
\end{equation}
while differentiating $g_E$ gives the \emph{conjugate} relations
\begin{equation}
\hat{m} = \alpha\,\partial_m g_E,\qquad \hat{R} = \alpha\,\partial_R g_E,\qquad \hat{q} = 2\alpha\,\partial_{\delta q} g_E,\qquad \delta\hat{q} = -2\alpha\,\partial_q g_E.
\end{equation}
Solving the conjugate equations as functions of the primal variables and substituting back, the free entropy reduces to an extremization over the physical order parameters alone,
\begin{equation}
\Phi_{t+1} = \operatorname*{extr}_{m,q,\delta q,R}\,\bigl[g_S(m,q,\delta q,R;\phi) + \alpha\,g_E(m,q,\delta q,R;\phi)\bigr],
\end{equation}
defining the map $\phi_t \mapsto \phi_{t+1} \equiv m^\star(\phi_t)$ quoted in the main text.

\subsubsection{Linear stability of the trivial fixed point}
\label{sec:replica_phi0_stability}

The transition line of Fig.~1(b) can be obtained by linearizing the map $\phi_{t+1} = f(\phi_t)$ around $\phi = 0$, without any numerical differentiation of the map. We keep the conventions of the previous subsection, suppressing the $t+1$ subscript on the student order parameters and writing $\phi \equiv \phi_t$.

The free entropy $\Phi_{t+1}$ inherits from the cluster relabeling $c \to -c$ a $\mathbb{Z}_2$ symmetry under which $(\phi, m, \hat{m})$ change sign while all remaining order parameters are unchanged. Along the saddle-point branch through the trivial fixed point $m = \hat{m} = 0$, the even order parameters $q$, $R$, $\delta q$ and their conjugates are therefore even functions of $\phi$ and are displaced only at order $\phi^2$, so that they may be replaced by their values at $\phi = 0$ (denoted by a subscript $0$). Only $m$ and $\hat{m}$ respond at linear order, and the stability parameter of the main text reads
\begin{equation}
r = \lim_{\phi \to 0} \frac{f(\phi)}{\phi} = \frac{1}{\sqrt{q_0}} \frac{\partial m}{\partial \phi} \bigg|_{\phi = 0}.
\end{equation}
Expanding the primal equation for $m$ and the conjugate equation $\hat{m} = \alpha \partial_m g_E$ to first order in $\phi$ gives the closed pair
\begin{equation}
m = \frac{\hat{m} + \hat{R}_0 \phi}{\lambda + \delta\hat{q}_0}, \qquad \hat{m} = \alpha \left( \chi_{mm} m + \chi_{m\phi} \phi \right),
\end{equation}
with the susceptibilities
\begin{equation}
\chi_{mm} = \frac{\partial^2 g_E}{\partial m^2}\bigg|_0, \qquad \chi_{m\phi} = \frac{\partial^2 g_E}{\partial m \partial \phi}\bigg|_0.
\end{equation}
Eliminating $\hat{m}$,
\begin{equation}
r = \frac{\hat{R}_0 + \alpha \chi_{m\phi}}{\sqrt{q_0} \left(\lambda + \delta\hat{q}_0 - \alpha \chi_{mm}\right)}.
\label{eq:r_criterion}
\end{equation}

Both susceptibilities reduce to one-dimensional Gaussian integrals, since $m$ enters $g_E$ only through the field $\omega_c = c m + \sigma \sqrt{q} z$ of the inner extremization defining $\mathcal{M}(c, z; y)$, while $\phi$ enters only through $A_c$. Writing $x^\star_y = \sigma \sqrt{\delta q_0} w^\star + \omega_c$ for the extremizer, the inner saddle-point equation reads
\begin{equation}
x^\star_y = \omega_c - \Gamma^2 \ell'(y, x^\star_y), \qquad \Gamma^2 = \sigma^2 \delta q_0,
\end{equation}
with primes denoting derivatives with respect to the second argument. Since $w^\star$ is stationary, the derivative of $\mathcal{M}$ with respect to $\omega_c$ acts only on the explicit dependence of the loss, and differentiating the inner saddle-point equation once more yields $\partial x^\star_y / \partial \omega_c = (1 + \Gamma^2 \ell'')^{-1}$, so that
\begin{equation}
\frac{\partial \mathcal{M}}{\partial \omega_c} = -\ell'\left(y, x^\star_y\right), \qquad \frac{\partial^2 \mathcal{M}}{\partial \omega_c^2} = -\frac{\ell''\left(y, x^\star_y\right)}{1 + \Gamma^2 \ell''\left(y, x^\star_y\right)}.
\end{equation}
The label symmetry $\ell(-1, x) = \ell(+1, -x)$ implies $\mathcal{M}(c, z; -1) = \mathcal{M}(-c, -z; +1)$, so that the $y = -1$ contributions can be folded onto $y = +1$ through the change of variable $z \to -z$.

For $\chi_{mm}$, at the trivial point $A_c = 0$ the sum over $c$ becomes trivial and only the second derivative of $\mathcal{M}$ contributes. Folding the two label branches,
\begin{equation}
\chi_{mm} = -2 \int \mathcal{D}z\, H(-B_0 z) \frac{\ell''(x^\star(z))}{1 + \Gamma^2 \ell''(x^\star(z))} \leq 0,
\end{equation}
where $x^\star(z) \equiv x^\star_{+1}$ evaluated at $\omega = \sigma \sqrt{q_0} z$, $B_0 = R_0 / \sqrt{q_0 - R_0^2}$, and for the logistic loss $\ell''(x) = h(x) h(-x)$ independently of the label. Since $\lambda + \delta\hat{q}_0 = 1/\delta q_0 > 0$, the sign of $\chi_{mm}$ guarantees that the denominator of Eq.~\eqref{eq:r_criterion} is strictly positive.

For $\chi_{m\phi}$, differentiating the $H$ weights with respect to $\phi$ produces the Gaussian density $G(x) = \e^{-x^2/2}/\sqrt{2\pi}$ with prefactor $\partial_\phi A_c = c a_0$, where $a_0 = \sqrt{q_0}/(\sigma\sqrt{q_0 - R_0^2})$, while the subsequent $m$ derivative acts on $\mathcal{M}$ through its first derivative. Folding the label branches as before and using $-\ell'(+1, x) = h(-x)$,
\begin{equation}
\chi_{m\phi} = 2 a_0 \int \mathcal{D}z\, G(B_0 z) h(-x^\star(z)) \geq 0.
\end{equation}
The product $\mathcal{D}z\, G(B_0 z)$ is an unnormalized Gaussian measure of variance $1/(1 + B_0^2) = (q_0 - R_0^2)/q_0$, and since $a_0 \sigma = \sqrt{1 + B_0^2}$ the prefactors combine into the compact form
\begin{equation}
\chi_{m\phi} = \sqrt{\frac{2}{\pi}} \frac{1}{\sigma} \left\langle h(-x^\star) \right\rangle_\omega, \qquad \omega \sim \mathcal{N}\left(0, \sigma^2 (q_0 - R_0^2)\right),
\end{equation}
where $x^\star$ solves the inner saddle-point equation at field $\omega$.

Finally, the criterion recovers the closed-form asymptote quoted in the End Matter. For $\lambda \to \infty$ the fields entering the inner saddle-point equation vanish, so $x^\star \to 0$, $\ell''(0) = 1/4$ and $h(0) = 1/2$, giving $\chi_{mm} \to -1/4$ and $\chi_{m\phi} \to 1/(\sigma\sqrt{2\pi})$, while the trivial saddle behaves as $\hat{R}_0 \to \alpha\sigma/\sqrt{2\pi}$, $\delta\hat{q}_0 \to \alpha\sigma^2/4$ and $\lambda^2 q_0 \to \alpha\sigma^2 \left(1/4 + \alpha/2\pi\right)$. Substituting into Eq.~\eqref{eq:r_criterion},
\begin{equation}
r \to r^\infty = \frac{1 + \sigma^2}{\sigma^2} \left(1 + \frac{\pi}{2\alpha}\right)^{-1/2},
\end{equation}
and the condition $r^\infty = 1$ yields the saturation value
\begin{equation}
\left(\sigma_c^\infty\right)^2 = \frac{2\alpha}{\pi} \left(1 + \sqrt{1 + \frac{\pi}{2\alpha}}\right).
\end{equation}

\subsection{Extension to label pooling} \label{sec:replica_pool}

We now extend the calculation to the setting where the student is trained against labels pooled from $T$ independent teachers. Each teacher $\W_t^{(\tau)}$, $\tau = 1, \dots, T$, contributes a provisional label on the same fresh batch of data, and the student minimizes a self-consistency loss against the resulting pooled label. The single-teacher calculation of the preceding section is recovered for $T=1$. We focus here only on the steps that genuinely differ from the $T=1$ case.

\subsubsection{Teacher partition function}

The teacher partition function now factorizes over the $T$ teachers, each constrained to lie on the same surface as in the single-teacher case (norm fixed to one, alignment with the centroid fixed to $\phi_t^{(\tau)}$):
\begin{equation}
Z_{t}=\int\prod_{\tau=1}^{T}\left[\dd\mu \left(\W_t^{(\tau)}\right)\delta \left(\|\W_t^{(\tau)}\|^{2}-N\right)\delta \left(\W_t^{(\tau)}\cdot \mu - N\phi_t^{(\tau)}\right)\right].
\end{equation}
Introducing Fourier representations of the delta constraints as in the single-teacher derivation produces one pair of conjugate variables $(\hat{q}_t^{(\tau)}, \hat{\phi}_t^{(\tau)})$ per teacher, and the integrand splits as $Z_t \sim \exp(N(g_I^{(t)} + g_S^{(t)})) = z_t^N$ with
\begin{align}
g_I^{(t)} &= -\sum_{\tau=1}^{T}\hat{q}_t^{(\tau)} q_t^{(\tau)} - \sum_{\tau=1}^{T}\hat{\phi}_t^{(\tau)}\phi_t^{(\tau)},\\
g_S^{(t)} &= \log\int\prod_{\tau=1}^{T}\frac{\dd w_t^{(\tau)}}{\sqrt{2\pi}}\,\e^{-\frac{1-2\hat{q}_t^{(\tau)}}{2}\bigl(w_t^{(\tau)}\bigr)^{2}+\hat{\phi}_t^{(\tau)} w_t^{(\tau)}}.
\end{align}
We will eventually fix all teachers to share the same alignment, $\phi_t^{(\tau)} = \phi_t$ for all $\tau$, in which case the saddle-point also imposes $\hat{q}_t^{(\tau)} = \hat{q}_t$ and $\hat{\phi}_t^{(\tau)} = \hat{\phi}_t$. The teacher entropic factor then collapses to $T$ identical contributions,
\begin{equation}
g_S^{(t)} =T\log\int\frac{\dd w_t}{\sqrt{2\pi}}\,\e^{-\frac{1-2\hat{q}_t}{2} w_t^{2}+\hat{\phi}_t w_t} = -\frac{T}{2}\log(1-2\hat{q}_t) +T\frac{\hat{\phi}_t^{2}}{2(1-2\hat{q}_t)},
\end{equation}
and the saddle-point relations $\hat{\phi}_t = -\phi_t/(\phi_t^{2}-1)$, $1-2\hat{q}_t = 1/(1-\phi_t^{2})$ from Eq.~\eqref{eq:teacher_saddle} carry over unchanged at the per-teacher level.

\subsubsection{Replicated student partition function}

Of the three contributions $g_I, g_S, g_E$ to the student free entropy, only the entropic term $G_S$ changes structurally: the student is now coupled to all $T$ teachers via student--teacher overlaps $R_t^{(\tau),a} = \sum_i \W_{t+1,i}^a \W_{t,i}^{(\tau)}/N$. The interaction term acquires one $\hat{R}_t^{(\tau),a} R_t^{(\tau),a}$ term per teacher, and the energetic term acquires the corresponding sum inside the Gaussian noise contribution. 
The entropic term in the presence of $T$ teachers now reads: 
\begin{align}
G_S = z_t^{-1}&\int\prod_{\tau}\frac{\dd w_t^{(\tau)}}{\sqrt{2\pi}}\,\e^{-\sum_{\tau}\frac{1-2\hat{q}_t^{(\tau)}}{2}(w_t^{(\tau)})^{2}+\sum_\tau \hat{\phi}_t^{(\tau)} w_t^{(\tau)}}\nonumber\\
&\times\int\prod_a \dd\mu \left(w_{t+1}^a\right)\e^{\sum_a \hat{m}_{t+1}^a w_{t+1}^a + \sum_{ab}\hat{q}_{t+1}^{ab}w_{t+1}^a w_{t+1}^b + \sum_{\tau}\hat{R}_t\sum_a w_{t+1}^a w_t^{(\tau)}}.
\end{align}
Hubbard--Stratonovich decoupling of $\hat{q}_{t+1}\bigl(\sum_a w_{t+1}^a\bigr)^2$ via the auxiliary variable $z$, and a change of variable for each teacher,
\begin{equation}
w_t^{(\tau)} \leftarrow \frac{\hat{\phi}_t^{(\tau)}}{1-2\hat{q}_t^{(\tau)}} + \frac{w_t^{(\tau)}}{\sqrt{1-2\hat{q}_t^{(\tau)}}}.
\end{equation}
At this stage, we may adopt the homogeneity assumption that we took in the computation of $Z_T$, canceling the latter in the denominator and simplifying the expression using the associated saddle-point relations. After this cancellation, the student is coupled to the teachers only through the pooled fluctuation field $\sum_\tau w_t^{(\tau)}$ and through the deterministic shift $T\hat{R}_t \phi_t$ inherited from the $\hat{\phi}_t/(1-2\hat{q}_t)$ piece of each teacher decomposition. Taking the limit $n \to 0$, applying the inverse-temperature rescalings, and performing the inner Gaussian integral over $w_{t+1}$ at its saddle point in the $\beta \to \infty$ limit, one obtains
\begin{equation}
g_S = \int\mathcal{D}z\int\prod_\tau \mathcal{D}w_t^{(\tau)}\,\frac{\left(\sqrt{\hat{q}_{t+1}} z + \dfrac{\hat{R}_t}{\sqrt{1-2\hat{q}_t}}\sum_\tau w_t^{(\tau)} +T\hat{R}_t \phi_t + \hat{m}_{t+1}\right)^{2}}{2(\lambda+\delta\hat{q}_{t+1})}.
\end{equation}
Performing the Gaussian integrals using $\int\mathcal{D}z\, z = 0$, $\int\mathcal{D}z\, z^{2} = 1$, and noting that the $T$ independent $w_t^{(\tau)}$ integrals contribute a combined variance of $T$ to the pooled fluctuation, the cross terms vanish and we are left with
\begin{equation}
g_S = \frac{\left(T\hat{R}_t \phi_t + \hat{m}_{t+1}\right)^{2} + \hat{q}_{t+1} + \dfrac{T\hat{R}_t^{2}}{1-2\hat{q}_t}}{2(\lambda+\delta\hat{q}_{t+1})}.
\end{equation}
Finally, substituting the teacher saddle-point relation $1-2\hat{q}_t = 1/(1-\phi_t^{2})$ from Eq.~\eqref{eq:teacher_saddle} yields the compact form
\begin{equation}
g_S = \frac{\left(T\hat{R}_t \phi_t + \hat{m}_{t+1}\right)^{2} + \hat{q}_{t+1} +T\hat{R}_t^{2}\,(1-\phi_t^{2})}{2(\lambda+\delta\hat{q}_{t+1})}.
\end{equation}
The $T=1$ case of the previous section is recovered as expected. 

\subsubsection{Saddle-point structure}

The remaining contributions $g_I$ and $g_E$ are obtained from the single-teacher expressions by replacing $\hat{R}_t \to T\hat{R}_t$ in $g_E$ (which inherits the same pooled coupling structure as $g_S$ after the homogeneity assumption), and leaving $g_I$ unchanged at the per-teacher level. Suppressing subscripts as before ($m \equiv m_{t+1}$, $q \equiv q_{t+1}$, $\delta q \equiv \delta q_{t+1}$, $R \equiv R_t$, $\phi \equiv \phi_t$), the free entropy reads
\begin{equation}
\Phi_{t+1} = \operatorname*{extr}_{m,q,\delta q,R}\,\bigl[g_S(m,q,\delta q,R;\phi,T) + \alpha\,g_E(m,q,\delta q,R;\phi,T)\bigr],
\end{equation}
and the iterated map $\phi_t \mapsto \phi_{t+1} = m^\star(\phi_t;T)$ now depends explicitly on the pool size $T$, giving the $T$-dependent phase boundary discussed in the main text.

\subsubsection{The $T \to \infty$ limit}

The saddle-point equations of the previous subsection admit a particularly clean asymptotic form as the pool size $T$ grows. Inspecting the structure, every explicit factor of $T$ in $g_S$ and in the modified $g_I = -[\hat{m}m + T\hat{R}R + \frac{1}{2}(\hat{q}\delta q - \delta\hat{q}q)]$ appears either as a drift contribution $T\hat{R}_t\phi_t$ or as a variance contribution $T\hat{R}_t^{2}(1-\phi_t^{2})$. The conjugate saddle-point equation
\begin{equation}
\hat{R}_t = \frac{\alpha}{T}\partial_R g_E
\end{equation}
forces $\hat{R}_t = O(1/T)$ as $T \to \infty$, since $g_E$ has no explicit $T$ dependence at fixed primal variables. This motivates the rescaling
\begin{equation}
\hat{R}_t = \frac{\hat{R}'_t}{T},\qquad \hat{R}'_t = O(1).
\end{equation}
Inserted into the finite-$T$ equations, this rescaling has two distinct consequences:
\begin{align}
T\hat{R}_t \phi_t &= \hat{R}'_t \phi_t \quad &&\text{(drift survives at $O(1)$)},\\
T\hat{R}_t^{2}\,(1-\phi_t^{2}) &= \frac{\hat{R}'^{2}_t (1-\phi_t^{2})}{T} \to 0 \quad &&\text{(variance suppressed at $O(1/T)$)}.
\end{align}

Substituting this rescaling into the finite-$T$ expressions and dropping the $O(1/T)$ remainder, the entropic term reduces to
\begin{equation}
g_S^{\infty} = \frac{\bigl(\hat{m}_{t+1} + \hat{R}'_t \phi_t\bigr)^{2} + \hat{q}_{t+1}}{2(\lambda + \delta\hat{q}_{t+1})},
\end{equation}
and the interaction term becomes
\begin{equation}
g_I^{\infty} = - \left[\hat{m}_{t+1}\,m_{t+1} + \hat{R}'_t R_t + \frac{1}{2} \left(\hat{q}_{t+1}\delta q_{t+1} - \delta\hat{q}_{t+1}q_{t+1}\right)\right].
\end{equation}
The energetic term $g_E$ retains its finite-$T$ form because, at the homogeneous-teacher saddle point, the pooled label channel is already determined by $(m_{t+1}, R_t, q_{t+1}, \delta q_{t+1}, \phi_t)$ alone; $T$ does not enter $g_E$ explicitly.

Extremization with respect to the conjugate variables yields the asymptotic primal equations
\begin{equation}
m_{t+1} = \frac{\hat{m}_{t+1} + \phi_t \hat{R}'_t}{\lambda+\delta\hat{q}_{t+1}},\qquad
R_t = \phi_t m_{t+1},\qquad
q_{t+1} = \frac{(\hat{m}_{t+1}+\phi_t\hat{R}'_t)^{2}+\hat{q}_{t+1}}{(\lambda+\delta\hat{q}_{t+1})^{2}},\qquad
\delta q_{t+1} = \frac{1}{\lambda+\delta\hat{q}_{t+1}},
\end{equation}
together with the conjugate equations
\begin{equation}
\hat{m}_{t+1} = \alpha\,\partial_{m_{t+1}} g_E,\qquad
\hat{R}'_t = \alpha\,\partial_{R_t} g_E,\qquad
\hat{q}_{t+1} = 2\alpha\,\partial_{\delta q_{t+1}} g_E,\qquad
\delta\hat{q}_{t+1} = -2\alpha\,\partial_{q_{t+1}} g_E.
\end{equation}
The conjugate equation for $\hat{R}'_t$ has lost its $1/T$ prefactor, reflecting the fact that $\hat{R}'_t$ is the natural $O(1)$ object in this limit. The collapse
\begin{equation}
R_t^{\star} = \phi_t m_{t+1}^{\star}
\end{equation}
is a clean signature of the asymptotic regime: in the infinite-pool limit, the student--teacher overlap becomes slaved to the product of the teacher and student alignments with the centroid, with no independent fluctuation degree of freedom along the teacher direction. Geometrically, the student is driven by the population-averaged signal direction and ignores the residual disorder carried by any individual teacher.

\subsubsection{Linear stability of the trivial fixed point with label pooling}

The analysis of the previous subsection extends directly to the pooled setting, because pooling leaves the energetic term unchanged. Each sample is labeled by a single teacher, and under the homogeneity assumption all teachers share the alignment $\phi$, so the joint statistics of a label and of the student pre-activation on a given sample are those of the single teacher case. The pool size enters $g_E$ only through the values of the order parameters, while its explicit dependence is confined to the entropic and interaction terms. The $\mathbb{Z}_2$ symmetry is likewise preserved, $g_S$ depending on the odd variables only through $(\hat{m} + T\hat{R}\phi)^2$, so that the even order parameters may again be replaced by their trivial saddle values. Expanding the primal relation for $m$ and the conjugate relation $\hat{m} = \alpha \partial_m g_E$ to first order in $\phi$,
\begin{equation}
m = \frac{\hat{m} + T\hat{R}_0 \phi}{\lambda + \delta\hat{q}_0}, \qquad \hat{m} = \alpha \left( \chi_{mm} m + \chi_{m\phi} \phi \right),
\end{equation}
with $\chi_{mm} \leq 0$ and $\chi_{m\phi} \geq 0$ given by the expressions of Sec.~\ref{sec:replica_phi0_stability}, with $B_0$ and $\Gamma^2$ now built from the pooled trivial saddle. Eliminating $\hat{m}$,
\begin{equation}
r(T) = \frac{T\hat{R}_0 + \alpha \chi_{m\phi}}{\sqrt{q_0} \left(\lambda + \delta\hat{q}_0 - \alpha \chi_{mm}\right)}.
\label{eq:rT_criterion}
\end{equation}
Since $T\hat{R}_0 = \alpha \partial_R g_E |_0$, no explicit factor of $T$ survives in the numerator, and the pool size acts on the ignition threshold only through the trivial saddle, in particular through
\begin{equation}
q_0 = \frac{\hat{q}_0 + T\hat{R}_0^2}{(\lambda + \delta\hat{q}_0)^2},
\end{equation}
whose second term, $(T\hat{R}_0)^2/T$, is suppressed as the pool growss, highlighting that pooling averages out the idiosyncratic component of the teachers' directions from the inherited weights, reducing $q_0$ at fixed signal transmission and thereby increasing $r(T)$. In the rescaled variables $\hat{R} = \hat{R}'/T$ of the previous subsection the numerator becomes $\hat{R}'_0 + \alpha \chi_{m\phi}$ and $q_0 \to \hat{q}_0/(\lambda + \delta\hat{q}_0)^2$, so that Eq.~\eqref{eq:rT_criterion} reduces at $T \to \infty$ to the linear stability criterion of the infinite pool theory. 

For $\lambda \to \infty$ the small field expansion of $g_E$ proceeds as in the single teacher case, giving $T\hat{R}_0 \to \alpha\sigma/\sqrt{2\pi}$, $\hat{q}_0 = \delta\hat{q}_0 \to \alpha\sigma^2/4$, $\chi_{mm} \to -1/4$ and $\chi_{m\phi} \to 1/(\sigma\sqrt{2\pi})$, all independent of $T$, while
\begin{equation}
\lambda^2 q_0 \to \alpha\sigma^2 \left(\frac{1}{4} + \frac{\alpha}{2\pi T}\right).
\end{equation}
Substituting into Eq.~\eqref{eq:rT_criterion},
\begin{equation}
r^\infty(T) = \frac{1+\sigma^2}{\sigma^2} \left(\frac{\pi}{2\alpha} + \frac{1}{T}\right)^{-1/2},
\end{equation}
which reduces to the single agent result at $T = 1$, increases monotonically with $T$, and saturates at $r^\infty(\infty) = r^\infty(1)\sqrt{1 + 2\alpha/\pi}$. The enhancement is bounded by the incoherent part of the sample noise, which survives pooling.

The condition $r^\infty(T) = 1$ gives the large $\lambda$ saturation of the ignition boundary,
\begin{equation}
\left(\sigma_c^\infty(T)\right)^2 = \left(\sqrt{\frac{\pi}{2\alpha} + \frac{1}{T}} - 1\right)^{-1},
\end{equation}
defined whenever $\pi/2\alpha + 1/T > 1$. For $\alpha < \pi/2$ this holds at every pool size, and the vertical asymptotes of the boundaries in the $(\sigma, \lambda)$ plane increase with $T$ and accumulate at $\left(\sigma_c^\infty(\infty)\right)^2 = \left(\sqrt{\frac{\pi}{2\alpha}} - 1\right)^{-1}$,
the maximal noise level at which collective ignition is possible at any regularization and pool size. For $\alpha > \pi/2$ the asymptote exists only for $T < T^\star = 2\alpha/(2\alpha - \pi)$, while for $T \geq T^\star$ one has $r^\infty(T) > 1$ at every noise level and the no learning phase no longer extends to large $\lambda$.

\bibliography{bibs}